\documentclass[twocolumn]{aastex631}
\usepackage{graphicx}
\usepackage{xspace} 
\usepackage{enumerate}
\usepackage{multirow}
\usepackage{threeparttable}

\newcommand{\OI}{[\textrm{O}~\textsc{i}]\xspace}
\newcommand{\OII}{[\textrm{O}~\textsc{ii}]\xspace}
\newcommand{\OIII}{[\textrm{O}~\textsc{iii}]\xspace}

\newcommand{\NII}{[\textrm{N}~\textsc{ii}]\xspace}
\newcommand{\SII}{[\textrm{S}~\textsc{ii}]\xspace}

\def \beq{\begin{equation}}
\def \eeq{\end{equation}}
\begin{document}

\title{Confirmation and Refutation of Lyman Continuum Leakers at $z\sim3$ with JWST NIRSpec/IFU}

\author[0009-0007-6655-366X]{Shengzhe Wang}
\affiliation{School of Astronomy and Space Science, University of Chinese Academy of Sciences (UCAS), Beijing 100049, China}
\affiliation{National Astronomical Observatories, Chinese Academy of Sciences, Beijing 100101, China}

\author[0000-0002-9373-3865]{Xin Wang}
\affiliation{School of Astronomy and Space Science, University of Chinese Academy of Sciences (UCAS), Beijing 100049, China}
\affiliation{National Astronomical Observatories, Chinese Academy of Sciences, Beijing 100101, China}
\affiliation{Institute for Frontiers in Astronomy and Astrophysics, Beijing Normal University, Beijing 102206, China}

\author[0009-0004-7133-9375]{Hang Zhou}
\affil{School of Astronomy and Space Science, University of Chinese Academy of Sciences (UCAS), Beijing 100049, China}

\author[0000-0002-0663-814X]{Yiming Yang}
\affil{National Astronomical Observatories, Chinese Academy of Sciences, Beijing 100101, China}
\affil{School of Astronomy and Space Science, University of Chinese Academy of Sciences (UCAS), Beijing 100049, China}

\author[0000-0001-7673-2257]{Zhiyuan Ji}
\affiliation{Steward Observatory, University of Arizona, 933 N. Cherry Avenue, Tucson, AZ 85721, USA}

\author[0009-0005-3823-9302]{Yuxuan Pang}
\affiliation{School of Astronomy and Space Science, University of Chinese Academy of Sciences (UCAS), Beijing 100049, China}

\author[0000-0002-9390-9672]{Chao-Wei Tsai}
\affil{National Astronomical Observatories, Chinese Academy of Sciences, Beijing 100101, China}
\affil{Institute for Frontiers in Astronomy and Astrophysics, Beijing Normal University,  Beijing 102206, China}
\affil{School of Astronomy and Space Science, University of Chinese Academy of Sciences (UCAS), Beijing 100049, China}

\author[0000-0002-7779-8677]{Akio K. Inoue}
\affiliation{Waseda Research Institute for Science and Engineering, Faculty of Science and Engineering, Waseda University, 3-4-1 Okubo, Shinjuku, Tokyo 169-8555, Japan}
\affiliation{Department of Physics, School of Advanced Science and Engineering, Faculty of Science and Engineering, Waseda University, 3-4-1 Okubo, Shinjuku, Tokyo 169-8555, Japan}

\author[0000-0001-5940-338X]{Mengtao Tang}
\affiliation{Tsung-Dao Lee Institute, Shanghai Jiao Tong University, 1 Lisuo Road, Shanghai 201210, People’s Republic of China}
\affiliation{School of Physics and Astronomy, Shanghai Jiao Tong University, 800 Dongchuan Road, Shanghai 200240, People’s Republic of China}
\affiliation{State Key Laboratory of Dark Matter Physics, Shanghai Jiao Tong University, 1 Lisuo Road, Shanghai 201210, People’s Republic of China}

\author[0000-0003-2804-0648]{Themiya Nanayakkara}
\affiliation{Centre for Astrophysics and Supercomputing, Swinburne University of Technology, Hawthorn, VIC 3122, Australia}

\author[0000-0002-3254-9044]{Karl Glazebrook}
\affiliation{Centre for Astrophysics and Supercomputing, Swinburne University of Technology, Hawthorn, VIC 3122, Australia}

\author[0000-0003-1718-6481]{Hu Zhan}
\affil{National Astronomical Observatories, Chinese Academy of Sciences, Beijing 100101, China}
\affil{Kavli Institute for Astronomy and Astrophysics, Peking University, Beijing 100871, China}

\author[0009-0007-5623-2475]{Pinjian Chen}
\affiliation{National Astronomical Observatories, Chinese Academy of Sciences, Beijing 100101, China}

\correspondingauthor{Xin Wang}
\email{xwang@ucas.ac.cn}

\begin{abstract}


Our understanding of the physical mechanisms and environments conducive to the escape of Lyman-continuum (LyC) radiation within the first 2 Gyr of cosmic history remains limited. Here we present a detailed analysis of JWST/NIRSpec medium-resolution IFU observations of two LyC-leaker candidates, LACES-94460 and LACES-104037 at $z = 3.1$, selected from deep HST/WFC3 F336W imaging and supported by ground-based spectroscopy. We first rule out LACES-94460 as a genuine LyC leaker, demonstrating that its apparent F336W signal originates from a nearby low-redshift interloper at $z = 1.6$, unambiguously identified through IFU spectroscopy. In contrast, for LACES-104037 we spectroscopically confirm bona fide LyC emission arising from a tidal-tail structure during the early stage of a galaxy merger, dubbed LACES104037–LyC. LACES104037–LyC exhibits extremely low rest-frame optical emission-line equivalent widths together with an exceptionally strong LyC flux. Within a picket-fence model framework, we reproduce its observed spectral and photometric properties with a young stellar population of age $\sim5$ Myr and a LyC escape fraction of $f_{\mathrm{esc}} \sim 99\%$. Our identification and detailed modeling of LACES104037–LyC provide one of the first compelling observational demonstrations for merger-driven LyC escape, indicating that galaxy mergers may represent an important and previously underappreciated contributor to the ionizing photon budget relevant for cosmic reionization. Furthermore, our analysis highlights the critical role of sub-kiloparsec resolution spectroscopy in securely identifying LyC leakers, removing contamination from closely projected low-redshift interlopers, and pinpointing the physical regions responsible for LyC leakage.

\end{abstract}


\section{Introduction}\label{intro}

The Epoch of Reionization \citep[EoR, $z\sim6$--11;][]{stark_galaxies_2016} marks the last major phase transition of the universe. Its central physical process is the gradual ionization of the intergalactic medium (IGM) by Lyman continuum (LyC) photons ($\lambda_{\rm rest}<912$\,\AA) produced by galaxies \citep{Dayal:2020ki}. 
Consequently, a central question in EoR studies is whether galaxies can produce and release sufficient LyC photons to drive cosmic reionization \citep{Ellis_2025,Robertson_2022ARA&A}.
Addressing this problem requires a clear understanding of the sources of LyC photons \citep{finkelstein_conditions_2019,naidu_rapid_2020,Jiang_2025}, the mechanisms by which they escape from dense interstellar medium (ISM) environments into the IGM \citep{carr_2025,Jaskot_lowz_review_2025, Gazagnes_2020,Flury_2025}, and the characteristic ionizing photon escape fraction ($f_{\mathrm{esc}}$;\citet{Gnedin_2008,Steidel_2001,Siana_2010,Izotov_2016,Izotov_2018_a,Izotov_2018_b}). 
Within $\Lambda$CDM-based reionization models, the $f_{\mathrm{esc}}$ is generally expected to lie in the range of $\sim10$–$20\%$ \citep{Inoue_2006,Robertson_2015,naidu_rapid_2020,Giovinazzo_2025}.

These questions must ultimately be answered observationally, motivating extensive efforts to identify galaxies with detectable Lyman continuum leakage (LyC leakers). Such samples are essential for placing empirical constraints on $f_{\mathrm{esc}}$ \citep{wang_lyman_2025, Steidel_2018,Ji_2020,Grazian_2017,Tanvir_2019}. Unfortunately, because the IGM transmission drops rapidly at $z>4$, direct observations of LyC leakers during the EoR are not feasible \citep{inoue_updated_2014}. As a result, attention has been shifted to lower-redshift analogs, where one can attempt to establish relationships between observable quantities and $f_{\mathrm{esc}}$ that may be extrapolated to the EoR \citep{Jaskot_lowz_review_2025}.

Thanks to the HST Cosmic Origins Spectrograph (COS), the largest LyC leaker sample to date has been assembled at $z\sim0.3$, consisting of $\sim70$ galaxies \citep[Low-redshift Lyman Continuum Survey; LzLCS;][]{Flury_2022_II,Izotov_2016,Le_Reste_2025_I}. 
This sample has enabled detailed investigations of the correlations between $f_{\mathrm{esc}}$ and a variety of observables, including the UV slope ($\beta_{UV}$) \citep{Chisholm_2022_UV_slope}, the \OIII/\OII\ ratio (O32) \citep{Pellegrini_2012,Jaskot_lowz_review_2025}, the effects of stellar feedback and ISM geometry \citep{Flury_2025}, and the Ly$\alpha$ emission-line profile \citep{Flury_2022_II}. These studies have revealed that young and extreme star formation activity is a necessary condition for LyC escape \citep{carr_2025,carr_effect_2024,kim_2023,Jaskot_lowz_review_2025}.
However, it must be noted that, due to the lack of high-resolution LyC band imaging observations for the low-$z$ sample, we cannot unambiguously determine which regions of the galaxies the escaping LyC photons originate from, nor the specific physical processes of the associated star-formation activity that trigger their escape \citep{Le_Reste_2025_III}.
Moreover, despite the presence of systems with very high $f_{\mathrm{esc}}$ \citep{Borthakur_2014,Flury_2022_I,izotov_detection_2016}, the average escape fraction in the LzLCS sample remains insufficient to account for the ionizing photon budget required for reionization \citep{Robertson_2022ARA&A}. 
In studies at high redshift, searches for LyC leakers have mainly been conducted within existing ultra-deep field datasets, rather than through systematic observations targeting sources with indirect evidence of LyC leakage \citep{Fletcher_2019,wang_lyman_2025,Zhu_2026,Seo_2026}. 
As a result, only a limited number of LyC leakers have been identified in the high-redshift Universe, while the large uncertainties associated with IGM absorption have further hindered definitive conclusions \citep{wang_lyman_2025,yuan_merging_2024,Steidel_2018}.

Galaxy mergers are widely regarded as a key physical process capable of enabling the escape of LyC photons \citep{Wang_Shengzhe_2025,zhu_lyman_2024,Weilbacher_2018, Le_Reste_2025_I}.
Through strong gravitational interactions, merging systems can simultaneously redistribute the interstellar medium---opening up low-column-density pathways---and induce bursts of star formation that substantially elevate the intrinsic LyC photon output \citep{Le_Reste_2025_III,yuan_merging_2024,zhu_lyman_2024,Faria_2025,Purkayastha_2022_GPs}.
In particular, merger-driven features such as tidal tails are capable of sustaining brief yet intense star-forming episodes \citep{Riess_2025,Weilbacher_2018}, which may rapidly deplete surrounding gas reservoirs and thereby facilitate the leakage of ionizing radiation \citep{Komarova_2024,Komarova_2021_MK71,Weilbacher_2018}.
Consistent with these observational findings, cosmological simulations at $z\sim5$–10 indicate that galaxy mergers can significantly increase $f_{\mathrm{esc}}$, even when employing simplified prescriptions for $f_{\mathrm{esc}}$ \citep{kostyuk_influence_2024}.

Observationally, the Ly$\alpha$ and Continuum Origins Survey (LaCOS, a subsample of LzLCS; \citep{Le_Reste_2025_I}) identified 22 LyC leakers, of which more than $41\%$ reside in merging systems \citep{Le_Reste_2025_III}. 
Although the parent LaCOS galaxy sample itself also contains a high merger fraction of $\sim48\%$ \citep{Le_Reste_2025_I}, these results suggest that galaxy interactions may facilitate the escape of LyC photons. At $z\sim3$, the LyC leaker candidates identified in the GOODS-S field by \citet{zhu_lyman_2024} frequently exhibit morphological signatures of galaxy interactions, as well as spatial offsets between the LyC emission and the main stellar component. Although contamination by low-$z$ interlopers cannot be fully excluded in these samples, these findings nevertheless indicate that mergers may provide favorable conditions for LyC escape.
A recently confirmed LyC leaker, J1244-LyC1, is a late-stage merger whose LyC emission is spatially distributed across multiple regions, further suggesting that the merger process can create multiple LyC escape sites \citep{Wang_Shengzhe_2025}.

Despite mounting observational evidence pointing to the importance of mergers in facilitating ionizing photon escape, only a limited number of studies have explicitly considered the role of mergers during the EoR, and none have reached firm conclusions \citep[see e.g.,][]{Mascia_2025}. 
In fact, the latest JWST observations indicate that the merger rate ($\mathcal{R}_{\mathrm{M}}$) is proportional to $(1+z)^2$ \citep{Calabr_2026}, implying that merger activity becomes increasingly prevalent toward higher redshift. This trend suggests that mergers may play an even more significant role during the EoR, where their impact on LyC escape could be substantially amplified.
This limitation primarily arises from the lack of high spatial resolution, multi-wavelength observations capable of directly linking merger-driven structures to $f_{\mathrm{esc}}$. Consequently, the identification and high-resolution characterization of merger-driven LyC leakers represent one of the most urgent observational priorities.

In this Letter, we present detailed reduction and analysis of JWST archival NIRSpec/IFU observations on two LyC-leaking candidates at $z\sim3.1$, LACES94460 and LACES104037. This paper is structured as follows. Section~\ref{sect:obs} describes the observations and data reduction. Section~\ref{sect:analysis} presents our analysis, including the identification of LyC leakage, emission-line diagnostics based on \texttt{PyNeb}, spectral energy distribution (SED) fitting, spatially resolved analyses of physical properties, and the determination of $f_{\mathrm{esc}}$ using stellar population models. Section~\ref{sect:discussion} discusses the implications of our results.
Throughout this work, we adopt a flat $\Lambda$CDM cosmology with $H_0 = 70~\mathrm{km~s^{-1}~Mpc^{-1}}$, $\Omega_{\mathrm{M}} = 0.3$, and $\Omega_{\Lambda} = 0.7$.

\section{Observations} \label{sect:obs}

LACES104037 (RA=334.2781, DEC=0.3592) and LACES94460 (RA=334.2834, DEC=0.3255) were identified as candidate LyC leakers from the LymAn Continuum Escape Survey \citep[LACES, HST-GO-14747, PI: Robertson,][]{Fletcher_2019}. Both sources have multi-band Subaru imaging as well as narrowband Ly$\alpha$ photometry, which initially confirmed them as Lyman-$\alpha$ emitters (LAEs) in the SSA22 protocluster field \citep{Hayashino_2004,Matsuda_2005,Yamada_2012}. 
Keck/MOSFIRE near-infrared spectroscopy detected strong rest-frame optical emission lines (e.g., \OII, \OIII, and H$\beta$), further confirming their membership in the SSA22 protocluster \citep{Nakajima_2020}.
Based on this comprehensive dataset, high spatial resolution HST imaging in the LyC band (F336W) enabled the identification of a sample of promising LyC-leaking galaxy candidates \citep{Fletcher_2019}. 
LACES104037 and LACES94460 were classified as members of the Silver sample owing to the significant spatial offsets between their F336W signals and their rest-UV continua.
In this work, we combine the latest JWST/NIRSpec IFU observations with existing archival data to perform a comprehensive investigation of LACES104037 and LACES94460.

\subsection{JWST IFU Data Reduction} \label{subsect:JWST_obs}

The data were obtained by JWST-GO-1827 (PI: Kakiichi). The NIRSpec IFU provides a $3'' \times 3''$ field of view (FoV) with an angular resolution of $0.1''$ \citep{Boker_2022}. Observations of these two sources were conducted with the medium-resolution grating (G235M/F170LP), using a 12-point dither pattern. This yielded spatially resolved spectra with a spectral resolution of $R \sim 1000$, covering the wavelength range of $1.66$--$3.17\,\mu$m. The spectra include key nebular emission lines such as H$\alpha$, H$\beta$, \OIII $\lambda4363$, the \OIII $\lambda\lambda4959,5007$ doublet and \SII doublet.

We performed a careful and customized reduction of the JWST/NIRSpec IFU data, primarily following the methodology presented in \citet{Fujimoto_2025}, while additionally adopting several non-standard processing techniques.
The full details of the data reduction procedure are described in Appendix~\ref{app:JWST_data_reduction}, where we also present a direct comparison between the data products before and after the refined processing (see Figure~\ref{fig:data_processing}). 
We emphasize that such a treatment is essential for our analysis, as instrumental effects are particularly pronounced in the regions of interest associated with LyC leakage. 
Only through cautious and careful data processing can the reliability of the subsequent scientific results be ensured.

In addition, due to the observational design, the WATA mode was not employed for auxiliary positional calibration, which renders precise astrometry a significant challenge. 
In Appendix~\ref{app:Astrometry}, we describe in detail our astrometric calibration strategy, through which we correct positional offsets of $\sim$0.2\arcsec\ for both LACES104037 and LACES94460.
We stress that accurate astrometric calibration is critically important for our study, as it constitutes a necessary prerequisite for distinguishing low-redshift interlopers from genuine LyC emission spatially offset from the main galaxy body.
We note that our astrometric solutions differ to some extent from those reported in \citet{Rivera-Thorsen_2025}, which we attribute to differences in the adopted calibration strategies.

\subsection{HST Data Reduction} \label{subsect:HST_obs}

High-resolution HST imaging of LACES104037 and LACES94460 was obtained with WFC3/UVIS and WFC3/IR as part of HST-GO-14747 \citep[PI: Robertson,][]{Fletcher_2019}. Both sources were observed in the F336W and F160W filters, probing rest-frame LyC (760--900\,\AA) and optical (3440--4100\,\AA) emission, respectively.


The HST imaging data were reprocessed following the approach described in \cite{Wang_Shengzhe_2025}, which enables more accurate relative astrometric calibration between individual exposures and further improves the overall image quality.
A detailed description of the updated reduction and alignment strategy will be presented in a forthcoming data release and initial science results paper (Wang et al., in preparation). 
The relative astrometric solution between different filters was derived using common compact sources and an affine transformation, performed directly in the native image coordinate systems. As a result, the alignment does not require the images to be resampled onto a common pixel grid, thereby preserving the full astrometric information available in the higher-resolution data.
The final mosaics have pixel scales of 0.03\arcsec\ and 0.06\arcsec\ for the F336W and F160W images, respectively.

Using the F160W image as the astrometric reference, we aligned the F336W image to the F160W frame. 
For LACES104037, we additionally aligned the JWST/NIRSpec IFU data to the F160W image (see Appendix~\ref{app:Astrometry}), ensuring a consistent relative astrometric registration among all three high spatial resolution datasets.
For LACES94460, JWST NIRCam F277W imaging observations are available \citep[JWST-GO-1869, PI: Schaerer,][]{Schaerer_2026}, whose wavelength coverage overlaps with that of the JWST NIRSpec IFU observations. Therefore, we directly performed astrometric alignment between the IFU data and the F277W image (see Appendix~\ref{app:Astrometry}).

\subsection{Ancillary Data}\label{subsect:Other_data}

For LACES104037, Keck/MOSFIRE $H$ and $K$ band spectroscopy covers a number of strong rest-frame optical emission lines (e.g., \OII, \OIII, and H$\beta$; \cite{Nakajima_2020}).
The emission-line measurements are taken from \cite{Nakajima_2020}. 


We performed SED fitting for both LACES104037 and its merging companion, incorporating the JWST spectroscopic data (Section~\ref{subsect:SED}). The primary photometric dataset follows that adopted in \citet{Fletcher_2019}. To separate the blended photometric contribution of the two components in the ground-based Subaru imaging, we remeasured the Subaru photometry using the high-resolution HST/F160W image as a prior and applied \textsc{TPHOT} \citep{Fletcher_2019,Yiming_2025}. 
This procedure enabled us to deblend LACES104037 and its merging companion and obtain independent photometric measurements for each component, which were then used in the subsequent SED fitting. 
The sum of the deblended fluxes is consistent with the total photometry reported by \citet{Fletcher_2019} (see Table~\ref{tab:photometry}), indicating that our remeasured photometry introduces no significant systematic offset.

\section{Analysis and Results} \label{sect:analysis}

\subsection{LyC Detection and Foreground Contamination} \label{subsect:LyC_D_contamination}

The high spatial resolution spectroscopic information provided by the JWST/NIRSpec IFU observations supplies a critical missing piece of observational evidence for assessing whether LACES104037 and LACES94460 are genuine LyC leakers.

\begin{figure*}[htbp]
    \centering
    \includegraphics[width=\textwidth]{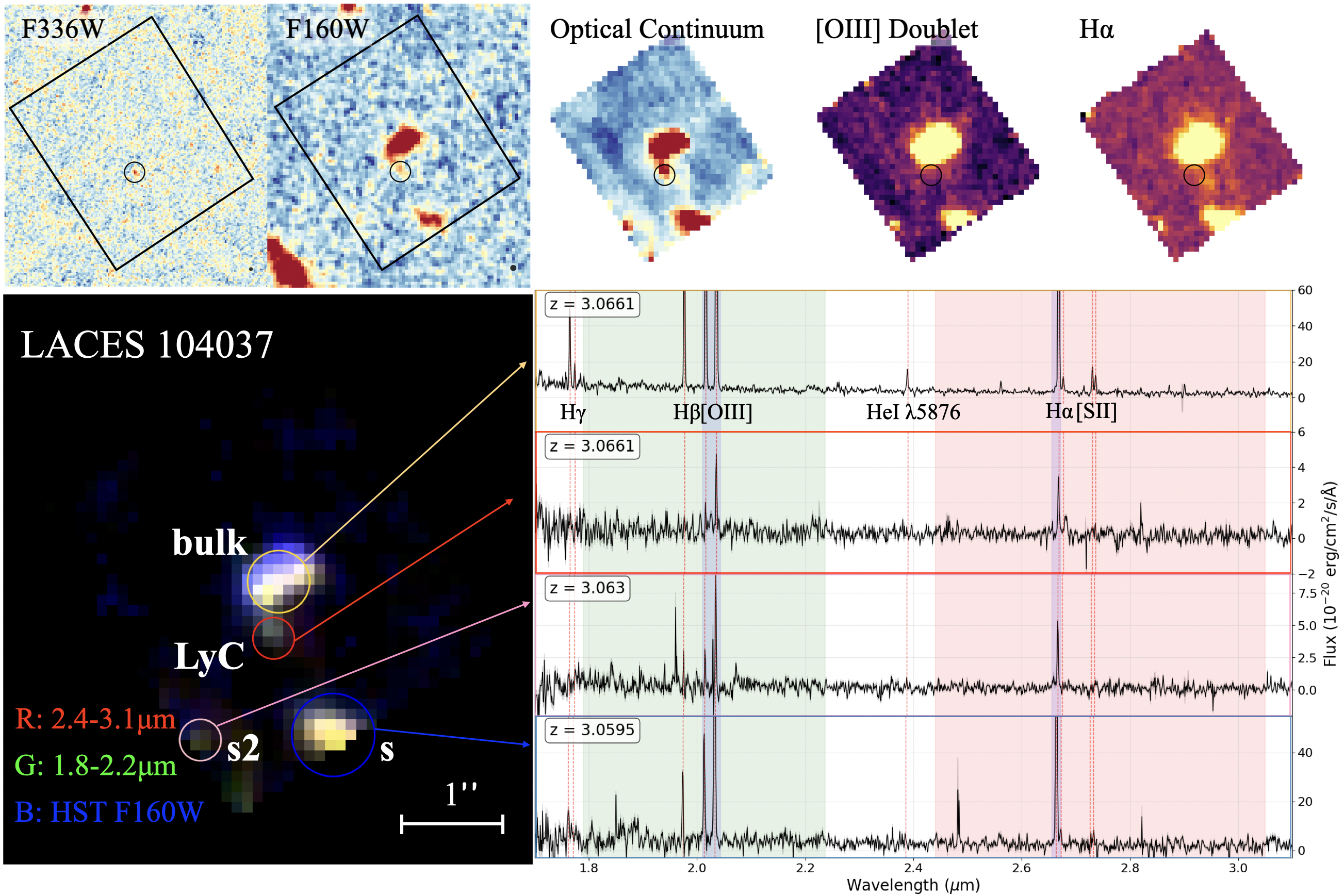}
 \caption{HST and JWST observations of LACES104037.
{\sc Top} panels: HST/WFC3 imaging obtained by the LACES program \citep[HST-GO-14747, PI: Robertson,][]{Fletcher_2019} and NIRSpec/IFU maps from the G235M/F170LP observations acquired by JWST-GO-1827 (PI: Kakiichi).
From left to right, we show the LyC-band (F336W), the NUV continuum (F160W), the optical continuum reconstructed from the IFU observations (with emission features masked), and the flux maps of \OIII and H$\alpha$. 
The black box marks the IFU FoV, while the black circle with a radius of 0.2\arcsec indicates the region of significant LyC leakage. 
The IFU maps are trimmed at the edges to exclude regions strongly affected by instrumental artifacts. 
{\sc Bottom} panels: three-color composite image and optimally extracted 1D spectra for the bulk of the system (LACES104037-bulk), the LyC-leaking region (LACES104037-LyC), and two mergers (LACES104037-s and LACES104037-s2), all spectroscopically confirmed at $z=3.06$.
The green and red shaded regions indicate the wavelength ranges corresponding to two of the color channels in the composite image, while the blue shaded region marks the wavelength coverage of the \OIII and H$\alpha$ emission-line maps.}
    \label{fig:LACES104037_data}
\end{figure*}

Figure~\ref{fig:LACES104037_data} presents the JWST/NIRSpec IFU observations of LACES104037. 
In the IFU FoV, we detect both LACES104037 and  two nearby companions. 
As those companions have no formal designation in NED, we refer to them as LACES104037s (RA=334.27789, DEC=0.35881) and LACES104037s2 (RA=334.27828, DEC=0.35881). 
For clarity, we separate the system into two components: the main galaxy body, where no LyC escape is detected (LACES104037-bulk; RA=334.27805, DEC=0.35923), and the spatially offset LyC-emitting region (LACES104037-LyC; RA=334.27805, DEC=0.35907).
We extract spectra for LACES104037-bulk, LACES104037-LyC, LACES104037s and LACES104037s2 from the IFU data cube (see Figure~\ref{fig:LACES104037_data}). 
The LyC-band (F336W) photometric flux of LACES104037-LyC is measured to be $7.6 \pm 0.9~\mathrm{nJy}$, at $\gtrsim$8 $\!\sigma$ significance.
All three spectra yield consistent spectroscopic redshifts of $z \simeq 3.066 \pm 0.001$. 
The projected separation between LACES104037 and LACES104037s is $\sim$12\,kpc. 
Notably, LACES104037-LyC exhibits a clear detection of optical continuum emission, indicating a significant stellar component and confirming that it is an integral part of the galaxy structure.
These results confirm that those objects form a merging system, with LyC leakage occurring in the region between the two interacting galaxies.

\begin{figure*}[htbp]
    \centering
    \includegraphics[width=\textwidth,trim=2cm 7cm 1cm 9.5cm,clip]{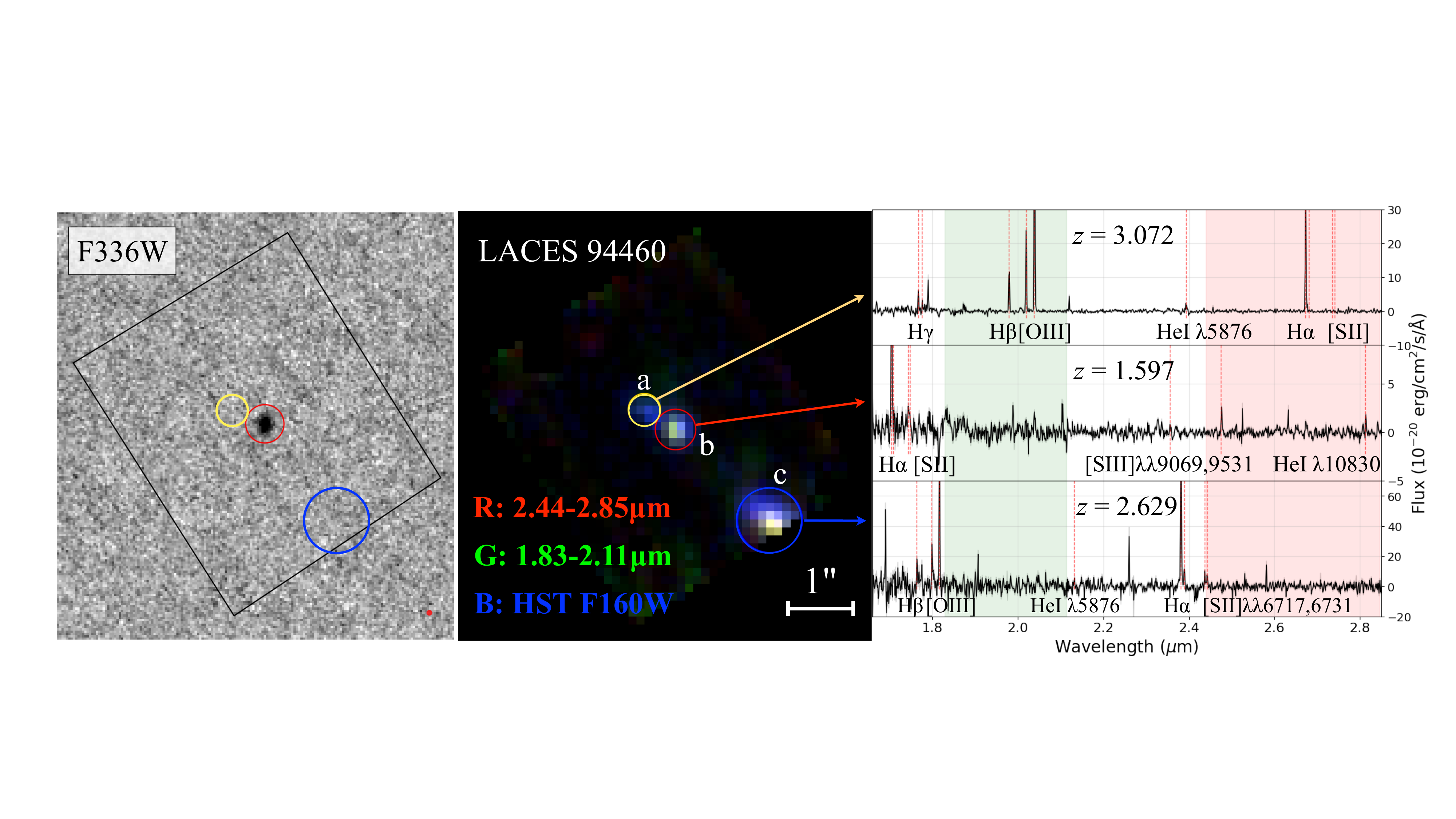}
    \caption{HST and JWST observations of LACES94460.
    {\sc Left}: HST WFC3/UVIS F336W image obtained by the LACES program \citep[HST-GO-14747, PI: Robertson,][]{Fletcher_2019}. The black box marks the FoV of the JWST NIRSpec/IFU G235M/F170LP observations acquired by JWST-GO-1827 (PI: Kakiichi).
    {\sc Middle}: three-color composite image with the HST F160W imaging as the blue channel and the JWST IFU data as the green and red channels.
    {\sc Right}: optimally extracted 1D spectra at $z=3.072$ (LACES94460-a), $z=1.597$ (LACES94460-b), and $z=2.629$ (LACES94460-c), with the corresponding emission lines highlighted.
    LACES94460 was previously identified as a silver LyC-leaking candidate at $z\sim3.1$ by \citet{Fletcher_2019}, based on its highly significant F336W detection spatially offset from UV continuum and its strong rest-optical oxygen lines confirmed by ground-based spectroscopy \citep[LACES94460-a,][]{Nakajima_2020}. However, our analysis demonstrates that the apparent F336W signal instead originates from a nearby low-redshift interloper, LACES94460-b at $z=1.6$, located within 0.5". This result highlights the critical importance of sub-kiloparsec resolution spectroscopy for reliably identifying LyC leakers at high redshifts and for removing contamination from closely projected low-redshift sources.
    }
    \label{fig:LACES94460_data}
\end{figure*}

Figure~\ref{fig:LACES94460_data} shows the JWST/NIRSpec IFU observations of LACES94460. 
In contrast to LACES104037, the IFU data for LACES94460 reveal a markedly different scenario. 
Within the IFU FoV, we identify three distinct sources LACES94460-a, LACES94460-b, and LACES94460-c at redshifts of $z=3.072$, $z=1.597$, and $z=2.629$, respectively.
With our refined astrometric calibration, we verify that the F336W emission spatially coincides with LACES94460-b at $z =  1.597$.
At this redshift, the F336W filter probes rest-frame UV emission at $\sim$1300\,\AA, rather than LyC radiation. 
No apparent signal of LACES94460-a at $z=3.072$ is seen in the F336W image.
We therefore rule out LACES94460 as a real LyC leaking galaxy at $z\sim3$.
Given the small angular separation of only $\sim$0.35\arcsec\ between the low-redshift interloper and LACES94460-a, such contamination can only be robustly identified through sub-kiloparsec resolution spectroscopy, such as NIRSpec/IFU or NIRISS wide-field slitless spectroscopy.

\subsection{Spectral Analysis}\label{subsect:Spec_Analysis}

\subsubsection{Emission Line Fluxes} \label{subsubsect:Emission_Lines}
Benefiting from the exceptional capabilities of JWST, we obtain the full rest-frame optical spectrum of LACES104037 covering 4200--7600\,\AA\ (see Figure~\ref{fig:LACES104037_data}). 
The spectrum exhibits a high signal-to-noise ratio (S/N) continuum together with several weak emission lines, including \OIII $\lambda4363$ and the \SII doublet.

To extract the emission-line spectrum, we perform a pixel-by-pixel continuum subtraction on the IFU data cube. 
After masking emission lines and outliers, the continuum in each spaxel is modeled using a fifth-order Chebyshev polynomial and subsequently subtracted. 
From the continuum-subtracted data cube, we extract an integrated emission-line spectrum of total LACES104037 using a circular aperture with a diameter of 1\arcsec, and measure the emission-line fluxes (see Table~\ref{tab:laces104037_lines}).

Keck/MOSFIRE $H$ and $K$ band spectroscopy detected the \OII and \OIII emission lines of LACES104037 \citep{Nakajima_2020}. 
Owing to seeing effects and slit losses in the Keck observations, the fluxes measured from the MOSFIRE spectra remain systematically lower than those obtained from the JWST/NIRSpec IFU data, even after applying slit-loss corrections. On average, the IFU-measured line fluxes are approximately 1.2 times higher than the corrected Keck values \citep[see Table~\ref{tab:laces104037_lines},][]{Nakajima_2020}.
We therefore adopt the \OIII $\lambda5007$/\OII doublet flux ratio measured from the Keck data, and combine it with the JWST \OIII line fluxes to infer the intrinsic \OII line flux (see Table~\ref{tab:laces104037_lines}).

We also measure the emission-line fluxes of LACES104037-LyC. 
The spectrum is extracted using a circular aperture with a diameter of 0.4\arcsec. 
As shown in Figure~\ref{fig:LACES104037_data}, only the \OIII and H$\alpha$ emission lines are significantly detected, while 3$\sigma$ upper limits are derived for all other measurable lines (see Table~\ref{tab:laces104037_lines}).

{
\tabletypesize{\scriptsize}
\begin{deluxetable}{lcc}
\tablecaption{Emission-line fluxes of LACES104037
\label{tab:laces104037_lines}}
\tablehead{
\colhead{Line} &
\colhead{LACES104037} &
\colhead{LACES104037-LyC} \\
&
\colhead{(erg s$^{-1}$ cm$^{-2}$)} &
\colhead{(erg s$^{-1}$ cm$^{-2}$)}
}
\startdata
H$\alpha$ &
$(8.529 \pm 0.423)\times10^{-17}$ &
$(9.892 \pm 0.941)\times10^{-19}$ \\
H$\beta$ &
$(2.736 \pm 0.596)\times10^{-17}$ &
$< 6.06\times10^{-19}$ \\
\OIII $\lambda5007$ &
$(1.800 \pm 0.008)\times10^{-16}$ &
$(1.519 \pm 0.169)\times10^{-18}$ \\
\OIII $\lambda4959$ &
$(5.920 \pm 0.054)\times10^{-17}$ &
$< 3.43\times10^{-19}$ \\
\NII $\lambda6583$ &
$(1.184 \pm 0.125)\times10^{-18}$ &
$< 1.20\times10^{-19}$ \\
\SII $\lambda6716$ &
$(3.741 \pm 0.198)\times10^{-18}$ &
$< 1.76\times10^{-19}$ \\
\SII $\lambda6731$ &
$(2.462 \pm 0.175)\times10^{-18}$ &
$< 2.23\times10^{-19}$ \\
\OI $\lambda6300$ &
$(9.201 \pm 2.329)\times10^{-19}$ &
$< 2.31\times10^{-19}$ \\
\OIII $\lambda4363$ &
$(2.434 \pm 0.467)\times10^{-18}$ &
$< 4.81\times10^{-19}$ \\
\OII $\lambda\lambda$3727,3730\tablenotemark{a} &
$(4.447 \pm 0.129)\times10^{-17}$ &
\ldots \\
\enddata
\tablecomments{
All reported uncertainties represent 1$\sigma$ errors. For the measurements on LACES104037-LyC with S/N$<3$, we report $3\sigma$ upper limits.
}
\tablenotetext{a}{Calculated using O32 obtained from ground observations \citep{Nakajima_2020}.}
\end{deluxetable}
}

\subsubsection{Emission Line Diagnostics} \label{subsubsect:Plasma_Diagnostics}

Based on the measured emission-line fluxes of H$\alpha$, H$\beta$, \OIII $\lambda5007$, \OIII $\lambda4363$ , and \SII doublet we performed emission line diagnostics using the \texttt{PyNeb} code \citep{Luridiana_pyneb_2015}.

To obtain accurate estimates of the dust attenuation, electron density ($n_{\mathrm{e}}$), and electron temperature ($T_{\mathrm{e}}$) for LACES104037, we follow the iterative approach described in \citet{Ueta_2022}. 
The calculation is performed iteratively until all three parameters converge.

Specifically, under the assumption of Case~B recombination, we adopt initial values of $n_{\mathrm{e}} = 100~\mathrm{cm^{-3}}$ and $T_{\mathrm{e}} = 10^{4}$\,K. 
An initial dust attenuation is derived from the H$\alpha$/H$\beta$ Balmer decrement and applied to correct the measured emission-line fluxes. 
We correct for reddening using the attenuation curve of \cite{Calzetti_2000}. 
The electron temperature is then determined from the \OIII $\lambda5007$ / \OIII $\lambda4363$ emission-line ratios, while the electron density is estimated from the \SII doublet. 
Using the updated values of $T_{\mathrm{e}}$ and $n_{\mathrm{e}}$, the dust attenuation is recalculated and the procedure is iterated until convergence is achieved.
Uncertainties on all derived quantities are propagated from the emission-line flux uncertainties using Monte Carlo simulations (see Table~\ref{tab:laces104037_phys_all}).
We obtain the following physical parameters for LACES104037: 
$E(B\!-\!V)=0.137^{+0.018}_{-0.020}$, 
$T_{\mathrm{e}}=13050^{+890}_{-941}\,\mathrm{K}$, and 
$n_{\mathrm{e}}=239^{+160}_{-132}\,\mathrm{cm^{-3}}$. 
These results indicate that LACES104037 exhibits non-negligible dust attenuation. 
The spatial non-uniformity of the dust extinction will be discussed in Section~\ref{subsubsect:Balmer_map}.

After correcting for dust attenuation, we derive the integrated star formation rate (SFR) of LACES104037 from the extinction-corrected H$\alpha$ luminosity (see Table~\ref{tab:laces104037_phys_all}) \citep{Kennicutt_1998}.
We note that metallicity variations can systematically affect the SFR calibration \citep{Bicker_2005,Inoue_2011}. According to \citet{Bicker_2005}, adopting a metallicity of $0.26\,Z_{\odot}$ for LACES104037 would lead to an SFR that is overestimated by a factor of $\sim1.8$ compared to the solar-metallicity calibration used in \citet{Kennicutt_1998}.
We therefore incorporate this effect into the systematic uncertainty of the derived SFR. 

\subsubsection{Metallicity and Ionization Parameter}\label{subsubsect:Metallicity}

With $T_{\mathrm{e}}$ and $n_{\mathrm{e}}$ determined, we combine the measured \OII and \OIII emission lines to compute the gas-phase oxygen abundances using \texttt{PyNeb}. 
Because the \OII $\lambda\lambda7320,7330$ auroral lines are not reliably detected, both the O$^{+}$ and O$^{++}$ abundances are derived using the electron temperature inferred from the \OIII lines and the electron density obtained from the \SII doublet.

As no He\,\textsc{ii} emission is detected in the spectrum, we do not apply an ionization correction factor (ICF) to account for a potential O$^{3+}$ contribution. 
Even if such a component were present, it would not significantly affect the total oxygen abundance of LACES104037. 
In practice, the O$^{++}$/H$^{+}$ abundance is independently derived from the \OIII $\lambda4363$ and \OIII $\lambda5007$ lines, and we adopt the mean of the two estimates. 
The O$^{+}$/H$^{+}$ abundance is derived from the \OII doublet. 

The total oxygen abundance is thus determined to be $12 + \log(\mathrm{O/H}) = 8.112 \pm 0.043$, with uncertainties propagated from the emission-line flux errors (see Table~\ref{tab:laces104037_lines}).

Adopting a global metallicity of $\sim$0.26\,$Z_{\odot}$ for LACES104037, we infer the integrated ionization parameter ($\log U $) from the O32 using the calibration of \citet{Kewley_logu_2002}, yielding $\log U = -2.41 \pm 0.10$.

\subsection{IFU Analysis}\label{subsect:IFU_Analysis}

\begin{figure*}[htbp]
    \centering
    \includegraphics[width=1\textwidth]{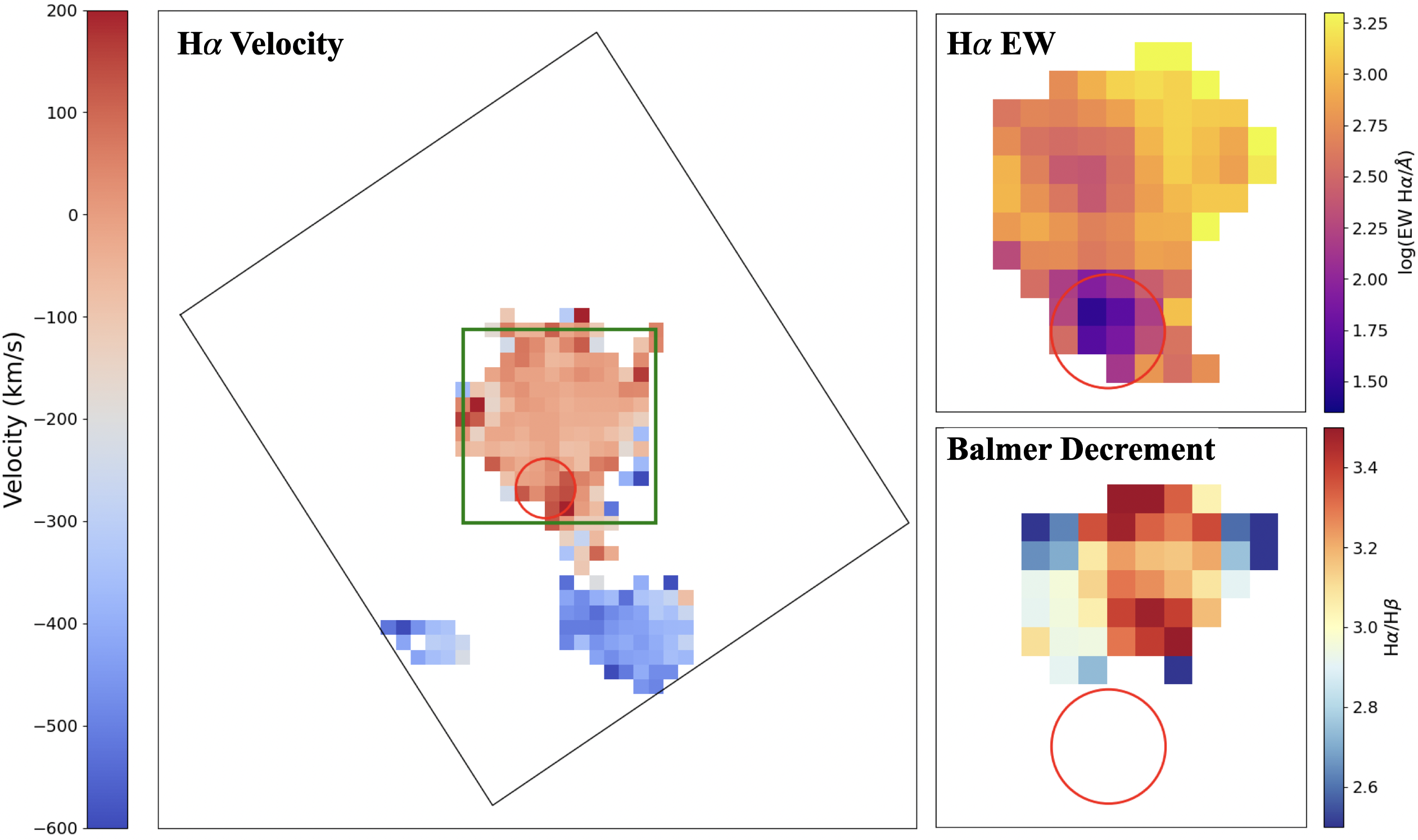}
    \caption{H$\alpha$ velocity, EW, and Balmer decrement maps of LACES104037.
    The {\sc left} panel shows the velocity distribution of the LACES104037 merger system, primarily consisting of three distinct components. 
    We adopt the redshift of LACES104037-bulk as the zero point and compute the velocity field of the entire system, which clearly reveals the relative velocity offsets among the three components.
    The black box indicates the JWST/NIRSpec IFU FoV, while the green box marks the spatial coverage of the panels shown on the right.
    The {\sc top right} panel presents the spatial distribution of H$\alpha$ EW in LACES104037. 
    The H$\alpha$ EW distribution within LACES104037-bulk is highly inhomogeneous, suggesting substantial differences in stellar population properties among individual star-forming clumps.
    The LACES104037–LyC region (marked by the red circle) exhibits an apparently much lower H$\alpha$ EW than the bulk component, which is caused by the significant escape of ionizing photons, breaking Case B recombination conditions.
    The {\sc bottom right} panel shows the Balmer decrement map.    
    Only pixels with H$\beta$ S/N $>$ 3 are included in the calculation. 
    The H$\alpha$ and H$\beta$ maps are smoothed with a $2 \times 2$ box filter prior to computing the Balmer decrement. 
    A pronounced spatial inhomogeneity is evident across the system. 
    We note that even after stacking all pixels within the LACES104037-LyC (marked by the red circle), the H$\beta$ S/N remains below the $3\sigma$ level.
    }
    \label{fig:maps}
\end{figure*}

The high spatial resolution spectroscopic information provided by the NIRSpec IFU enables us to construct spatially resolved maps of key physical properties. 
In this work, we focus on the velocity field, the H$\alpha$ equivalent width (EW) distribution, and the H$\alpha$/H$\beta$ map of the LACES104037 merging system (Figure~\ref{fig:maps}).

\subsubsection{Velocity field}
We derive the line-of-sight velocity field from the H$\alpha$ emission line, adopting the systemic redshift of LACES104037 as the zero-velocity reference. 
As shown in Figure~\ref{fig:maps}, we restrict the analysis to regions with H$\alpha$ S/N $> 3$. 
A clear velocity offset of $\sim$480\,km\,s$^{-1}$ is observed between LACES104037 and LACES104037s. 
Given the spectral resolution of the NIRSpec G235M grating ($R \sim 1000$, corresponding to a velocity resolution of $\sim$300\,km\,s$^{-1}$), this velocity difference is robustly resolved. 
The velocity map further reveals diffuse ionized gas bridging the two galaxies, consistent with an interacting and merging system.

\subsubsection{Emission-Line EW Map}
We construct the H$\alpha$ EW map using spaxels with H$\alpha$ integrated flux S/N $> 3$.
The map of the integrated stellar continuum is obtained by stacking the line-free regions of the IFU data cube (see Appendix~\ref{app:JWST_data_reduction} and Figure~\ref{fig:LACES104037_data}). 
Combining this continuum map with the H$\alpha$ line flux map allows us to compute the spatially resolved EW distribution. 
Under fiducial Case B recombination conditions, the EW of hydrogen recombination lines is closely linked to the duration of recent starburst activity, with larger EWs indicating younger stellar populations assuming constant star formation history (SFH) \citep[see e.g.,][]{Zanella_2015}. 
The resulting H$\alpha$ EW map is highly non-uniform and exhibits spatial gradients.  
Notably, the H$\alpha$ EW in the LACES104037-LyC is lower than the LACES104037-bulk average by approximately an order of magnitude (Figure~\ref{fig:maps}).
We discuss the implications of this result in detail in Section~\ref{subsect:SFH_Model}.

\subsubsection{Balmer Decrement Map}\label{subsubsect:Balmer_map}
The H$\alpha$/H$\beta$ map is constructed using spaxels with H$\beta$ S/N $> 3$ and traces the spatial variation of dust attenuation across the system (see Figure~\ref{fig:maps}). 
Under the assumption of the intrinsic Case~B recombination line ratio, lower H$\alpha$/H$\beta$ values correspond to lower dust attenuation. 
LACES104037 exhibits pronounced spatial variations in dust attenuation. 
In particular, within the LACES104037-LyC, no significant H$\beta$ emission is detected even when stacking the surrounding $3 \times 3$ spaxels.
The resulting $3\sigma$ lower limit on the H$\alpha$/H$\beta$ ratio is $\sim1.63$, which is consistent with the dust-free, standard Case~B value of 2.86 \citep{Osterbrock_2006_ISM}. 
Given the extremely strong LyC emission observed in LACES104037-LyC, we therefore consider the assumption that LACES104037-LyC is effectively dust-free to be reasonable.

\subsection{Spectral Energy Distribution and UV slope}\label{subsect:SED}

\begin{figure*}[htbp]
    \centering
    \includegraphics[width=1\textwidth]{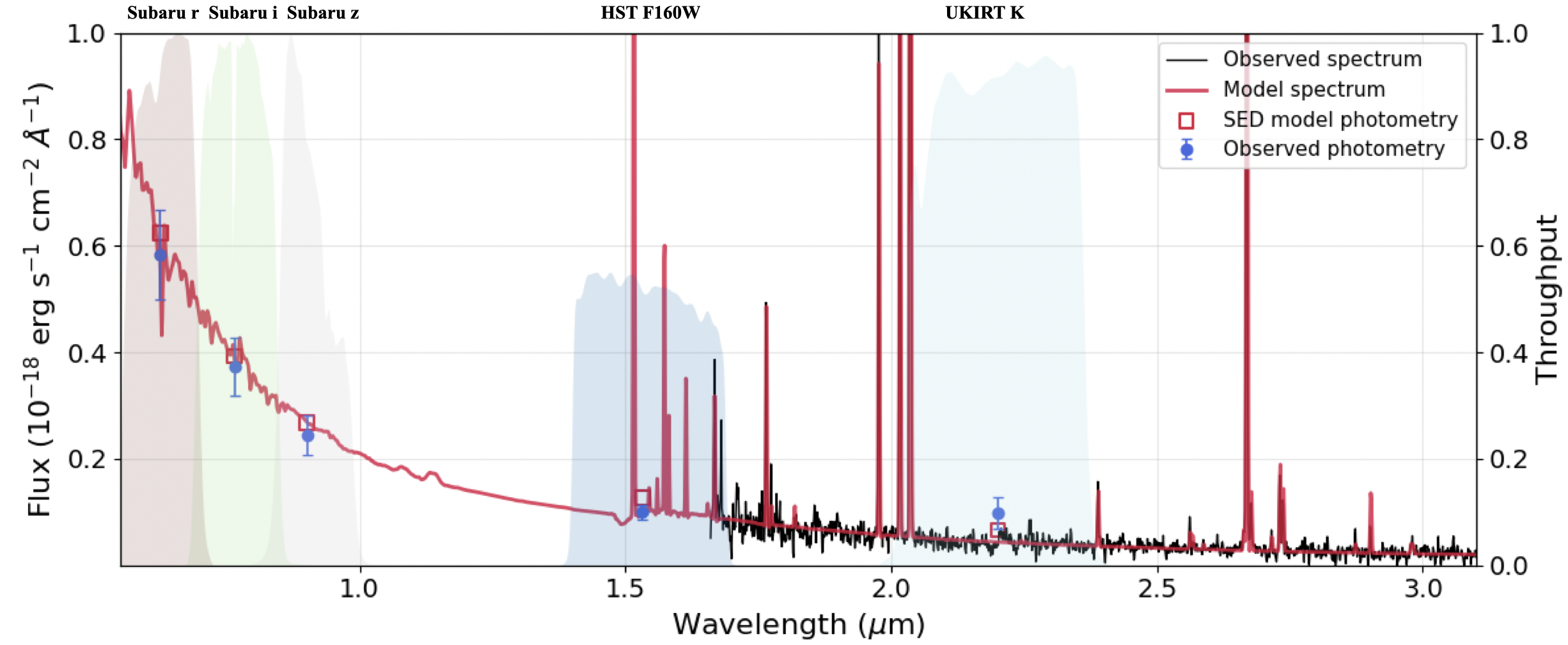}
    \caption{Best-fit SED model of LACES104037 derived using \texttt{PROSPECTOR}, fitted to the existing broad-band photometry and IFU spectroscopy. Our spectral fitting results are shown in Table~\ref{tab:laces104037_phys_all}. 
    The best-fit SED model yields a reduced chi-square of $\chi^2 = 2.93$.
    LACES104037 is characterized by a very young stellar population, with its SED dominated by a moderately evolved component of age $\sim16$ Myr, while an additional recent starburst within the past $\sim1$ Myr is required to reproduce the observed spectroscopic and photometric properties.
    } 
    \label{fig:SED_result}
\end{figure*}

We use the \texttt{Prospector} stellar population synthesis code to simultaneously fit the JWST/NIRSpec IFU spectrum of LACES104037 together with all available photometric data \citep{Johnson_prospector_2021}. 
The photometric measurements are primarily taken from \citet{Fletcher_2019}, including Subaru $R$, $i'$, and $z'$ bands, HST F160W, and the UKIRT $K$ band. 
We do not include the CFHT $u$ band or the Subaru $B$ and $V$ bands in the SED fitting, because their wavelength coverage lies near or blueward of Ly$\alpha$. 
Since the SED fitting procedure does not adequately model the complex physics associated with Ly$\alpha$ emission and absorption, including these bands would introduce additional systematic uncertainties and bias the inferred physical properties of LACES104037-bulk \citep{Sandberg_2015}. We therefore exclude these bands from the SED fitting.
In addition, the two Spitzer Space Telescope bands are close to the detection limit and are likewise not used in the SED fitting.
We note that the UKIRT $K$ band photometry lies within the wavelength coverage of our IFU observations, and the absolute flux calibration between the two observations is consistent within the $1\sigma$ uncertainties (see Figure~\ref{fig:SED_result}). Therefore, no renormalization of the IFU spectra to match the broadband photometry is applied.


In Section~\ref{subsect:Spec_Analysis}, we discuss the constraints on the global physical properties of LACES104037 provided by the available spectroscopic observations, including metallicity, dust attenuation, and the ionization parameter $\log U$. 
Because the uncertainties in these spectroscopically derived parameters are smaller than the uncertainties inherent to the SED fitting itself, we fix these three parameters during the Prospector fitting.


We adopt a double delayed-$\tau$ SFH model in the SED fitting \citep{Suess_2022_SFH}. 
Compared to the single-component delayed-$\tau$ models, which are among the most commonly used SFH prescriptions, the double delayed-$\tau$ model provides a more flexible and physically motivated description of the star formation history of merging systems \citep{Tissera_2002}. 
Such systems are naturally characterized by two relatively distinct phases of star formation: an extended pre-merger star formation episode and a merger-triggered starburst.

We constrain the onset time of the long-timescale pre-merger SFH ($t_{\mathrm{burst,old}}$) to be less than 1 Gyr, and limit the onset time of the short-timescale merger-triggered starburst ($t_{\mathrm{burst,young}}$) to be less than 500 Myr.
No explicit constraints are imposed on the e-folding timescales $\tau_{\mathrm{old}}$ and $\tau_{\mathrm{young}}$, for which we adopt uniform priors.
As shown in Figure~\ref{fig:picket-fence Model}, the emission-lines EWs measured for LACES104037-bulk are all above 200 \AA, with an average value of $\sim$600 \AA. 
Based on the empirical relation between EW and star formation age presented by \cite{Zanella_2015}, such large EWs imply a very young stellar population. Although the observed EWs may be reduced relative to the intrinsic values assumed in that relation because of ionizing photon escape, this effect would make the inferred stellar population age older. Therefore, the inferred age should be regarded as a conservative upper limit, and we consider the adopted upper limits on the SFH timescales in our SED modeling to be sufficiently conservative.

The best-fitting results are shown in Figure~\ref{fig:SED_result}, the derived SED-fitting parameters are summarized in Table~\ref{tab:laces104037_phys_all}.
Combining the dust attenuation derived from the Balmer decrement, $E(B\!-\!V)\sim0.14$, with the stellar mass obtained from the SED fitting, $\sim10^{8.8}\,M_\odot$, we infer that LACES104037 is a starburst dwarf galaxy with significant dust attenuation.

We also perform SED fitting for LACES104037s (see Appendix~\ref{app:SED_Results}). 
Emission-line diagnostics indicate a relatively high dust attenuation, with $E(B\!-\!V) \simeq 0.56$.
Combined with the SED constraints, we find that LACES104037s is a heavily obscured galaxy with a stellar mass of $\sim10^{9}\,M_\odot$, forming a major-merger system together with LACES104037.

For LACES94460, ground-based observations are unable to resolve the source. Therefore, only the F160W photometry and the IFU spectroscopic data are used for the SED fitting (see Appendix~\ref{app:SED_Results}. Based on the resulting SED model, we infer that LACES94460 is a starburst dwarf galaxy with a stellar mass of approximately $10^{7.6}\,M_{\odot}$.

Based on the Subaru $V$, $R$, $i’$, and $z’$ broadband photometric observations \citep{Fletcher_2019}, we measured a UV continuum slope of $\beta_{\rm UV} = -2.11 \pm 0.20$ for LACES104037 (see Table~\ref{tab:laces104037_phys_all}). We then corrected the measured $\beta_{\rm UV}$ for dust attenuation using the extinction derived for the integrated galaxy \citep[see Section~\ref{subsubsect:Plasma_Diagnostics}]{Calzetti_2000}, yielding a dust-corrected UV slope of $\beta_{\rm UV} = -2.59 \pm 0.20$. Throughout this work, we adopt the \citet{Calzetti_2000} attenuation law for consistency. We note, however, that lower-metallicity galaxies may instead follow a steeper attenuation curve (e.g., \citealt{Shivaei_2020}), which would likely imply an intrinsically bluer $\beta_{\rm UV}$ than the value derived here. Furthermore, the total luminosity of LACES104037-LyC is only about one-fortieth that of LACES104037-bulk. The measured $\beta_{\rm UV}$ is therefore expected to be overwhelmingly dominated by the light from LACES104037-bulk and can be regarded as representative of the bulk component.

\begin{deluxetable}{lc}
\tablewidth{0.8\textwidth}
\tablecaption{Physical properties of LACES104037 \label{tab:laces104037_phys_all}}
\tablehead{
\colhead{Parameter} &
\colhead{Value}
}
\startdata
$T_{\mathrm{e}}$\tablenotemark{a} (K) &
$13050^{+890}_{-941}$ \\
$n_{\mathrm{e}}$\tablenotemark{a} (cm$^{-3}$) &
$239^{+160}_{-132}$ \\
$E(B\!-\!V)$\tablenotemark{a} &
$0.137^{+0.018}_{-0.020}$ \\
$12+\log(\mathrm{O/H})$\tablenotemark{a} &
$8.112^{+0.043}_{-0.043}$ \\
$\log(U)$\tablenotemark{b} &
$-2.41^{+0.10}_{-0.10}$ \\
SFR\tablenotemark{c} ($M_{\odot}\,\mathrm{yr}^{-1}$) &
$72 \pm 36$ \\
$\log_{10}(M_{\star}/M_{\odot})$\tablenotemark{d} &
$8.83 \pm 0.01$ \\
$t_{\mathrm{burst,old}}$\tablenotemark{d} (Myr) &
$15.86 \pm 0.02$ \\
$\tau_{\mathrm{old}}$\tablenotemark{d} (Gyr) &
$4.67 \pm 0.41$ \\
$t_{\mathrm{burst,young}}$\tablenotemark{d} (Myr) &
$< 1$ \\
$\beta_{UV}$\tablenotemark{e} &
$-2.11 \pm 0.20$ \\
\enddata
\tablecomments{All reported uncertainties represent 1$\sigma$ errors.}
\tablenotetext{a}{Derived using \texttt{PyNeb} based on emission-line diagnostics.}
\tablenotetext{b}{Estimated following \cite{Kewley_logu_2002}.}
\tablenotetext{c}{Estimated following \cite{Kennicutt_1998}.}
\tablenotetext{d}{Derived from SED fitting. We note that $t_{\mathrm{burst,young}}$ reaches the lower bound (1 Myr) allowed in the fitting procedure, indicating that the true value is likely younger than 1 Myr. Consequently, $\tau_{\mathrm{young}}$ is not well constrained and is therefore not reported.}
\tablenotetext{e}{Before dust-correction.}
\end{deluxetable}

\subsection{LyC escape fraction}

As discussed in \citet{Wang_Shengzhe_2025}, the estimation of $f_{\mathrm{esc}}$ is the central issue in the study of LyC leakers, and multiple approaches have been proposed. \citet{Fletcher_2019} adopted an SED-based method to estimate the global $f_{\mathrm{esc}}$ of LACES104037, obtaining $f_{\rm esc} \sim 13\%$. However, based on JWST IFU observations, we confirm that LACES104037-bulk does not exhibit significant ionizing photon escape, and that the dominant source of escaping ionizing photons originates from LACES104037-LyC. Therefore, it is essential to estimate $f_{\mathrm{esc}}$ specifically for LACES104037-LyC.

However, implementing the approach of \citet{Fletcher_2019} is highly challenging, as it requires multi-band, high spatial resolution imaging, particularly with rest-UV coverage. For LACES104037-LyC, only rest-frame LyC band (F336W), optical $U$ band (F160W), and optical IFU spectroscopy (F170LP) observations are available. This prevents us from placing effective constraints on its UV band emission. 
Another commonly adopted approach is to estimate $f_{\mathrm{esc}}$ from an energy budget perspective, which relies on the spectral SED within the LyC band. When stellar population models are involved, such estimates suffer from substantial uncertainties \citep{Rivera-Thorsen_2025}.

Considering these difficulties, we develop a new method to estimate $f_{\mathrm{esc}}$ for LACES104037-LyC.
We utilize the \texttt{Starburst99} stellar population synthesis code \citep{Leitherer_1999_starburst} and the \texttt{MAPPINGS}  \citep{Sutherland_2018} photoionization code to self-consistently forward model the observed LyC flux and rest-optical spectra of LACES104037-LyC, assuming constant SFH and the picket-fence model. A detailed description of our methodology will be presented in \citet{Zhou.2026}.

\subsubsection{Picket-fence Model}

To estimate the local $f_{\mathrm{esc}}$, we adopt the picket-fence model under a dust-free assumption. This model assumes a mixture of density-bounded and ionization-bounded nebular regions along the line of sight, such that $f_{\mathrm{esc}}$ is directly linked to the covering fraction ($C_f$) of optically thick gas, with $f_{\rm esc} = 1 - C_f$ \citep{Jaskot_lowz_review_2025, Giovinazzo_2025, marques-chaves_extreme_2022}. In practice, the stellar population is divided into two components: the obscured component, for which $f_{\rm esc}=0$, and the unobscured component, for which $f_{\rm esc}=1$.
This model further assumes negligible dust attenuation along the LyC escape channels, which is expected to be a reasonable approximation for LACES104037-LyC.

Operationally, we first generate constant-SFH stellar population spectra of different ages using \texttt{Starburst99}.
The obscured stellar population component is given by the \texttt{Starburst99} stellar population spectra multiplied by $C_f$, and the resulting spectra are then input into \texttt{MAPPINGS} to generate the corresponding nebular emission.
The final spectrum then corresponds to the combination of the unobscured component (i.e., the \texttt{Starburst99} stellar population spectra multiplied by $f_{\rm esc}=1-C_f$) and the emergent obscured component whose ionizing radiation has been fully absorbed following Case~B producing nebular continuum and lines.
Although the actual ISM structure is inevitably more complex than that described by the picket-fence model \citep{Jaskot_lowz_review_2025, Jaskot_2019}, especially in extreme LyC leakers where fully optically thick ionization-bounded nebulae may not exist \citep{Gazagnes_2020}, this simplified framework provides an intuitive understanding of observable signatures of ionizing photon escape. In particular, in cases of extreme $f_{\mathrm{esc}}$, differences between various model assumptions are significantly reduced \citep{Jaskot_lowz_review_2025}.

\subsubsection{Stellar and Nebular Modelling}

We assume a constant SFH and generate stellar population spectra at ages of 3, 5, 10, 20, and 50 Myr using \texttt{Starburst99}, adopting the default broken power-law IMF with slopes of $\alpha_1=-1.30$ over the mass range $0.1$--$0.5\,M_{\odot}$ and $\alpha_2=-2.3$ over $0.5$--$120\,M_{\odot}$ \citep{Kroupa_2001_IMF, Leitherer_1999_starburst}. The stellar metallicity is set to $0.2\,Z_{\odot}$, and a combination of PAULDRACH/HILLIER stellar atmosphere models is adopted \citep{Pauldrach_2001,Hillier_1998}. 
We note that the specific choice of stellar population synthesis model mainly affects the predicted ionizing spectrum. In the present calculations, however, we treat $f_{\mathrm{esc}}$ as a free parameter and focus on the radiative consequences of a given $f_{\mathrm{esc}}$, rather than on the physical processes that establish it through stellar feedback. Consequently, although binary stellar population models (e.g., BPASS, \citet{Byrne_2022}) may influence the evolution of the ISM and the resulting $f_{\mathrm{esc}}$ (e.g., \citealt{Ma_2016,Kimm_2019,Rosdahl_2018}), we do not expect the qualitative trends presented here to depend sensitively on the adopted stellar population synthesis model \citep{Stanway_2016}.

The resulting stellar population spectra are multiplied by the factor $C_f$ to represent the fraction obscured by optically thick gas, and then input into \texttt{MAPPINGS}. The nebular parameters are fixed to match those of LACES104037-bulk, with a metallicity of $0.26\,Z_{\odot}$ , $\log U=-2.41$, $T_e=13\,000$~K, and $n_e=250~\mathrm{cm^{-3}}$. These parameters uniquely determine the resulting nebular emission-line ratios (Section~\ref{subsect:Spec_Analysis}; see Table~\ref{tab:laces104037_phys_all}).

The final composite spectra are obtained by combining the stellar population spectra from \texttt{Starburst99} with the nebular spectra from \texttt{MAPPINGS}. Since the obscured stellar population component contributes zero escaping ionizing photons, the LyC region of the composite spectra is multiplied by $f_{\mathrm{esc}}$ to account for the effect of the picket-fence model.

\begin{figure}[htbp]
    \centering
    \includegraphics[width=0.5\textwidth]{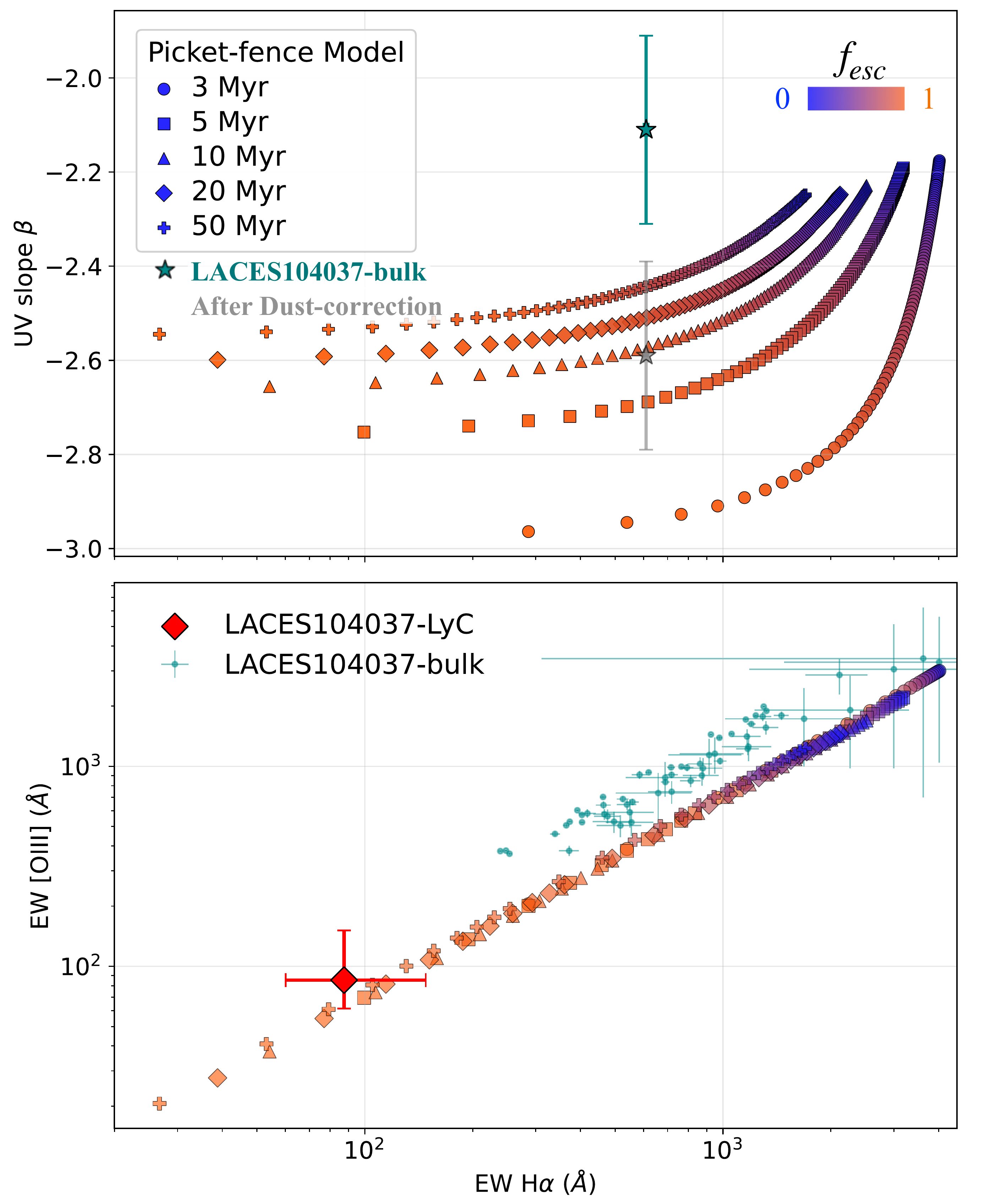}
    \caption{Relation between $f_{\mathrm{esc}}$ and observable quantities (emission-line EW and $\beta_{UV}$) in the picket-fence model. {\sc Top} panel shows the variation of $f_{\mathrm{esc}}$ in the $\beta_{UV}$--EW(H$\alpha$) parameter space for different stellar ages, with the observational result of LACES104037-bulk overplotted. {\sc Bottom} panel illustrates that $f_{\mathrm{esc}}$ and stellar age are fully degenerate when only emission-line EWs are considered. H$\alpha$ and \OIII are the two emission lines observable for LACES104037-LyC, and in the EW-based parameter space, variations in age and $f_{\mathrm{esc}}$ trace the same evolutionary locus. The pixel-by-pixel measurements of LACES104037-bulk (with emission-line S/N $>10$ per pixel) and the observational result of LACES104037-LyC are both overplotted, revealing a pronounced gap between the two. We note that our models include only a young stellar population, while the observed optical continuum may contain a contribution from an older stellar population (see the stellar population parameters derived from SED fitting in Table~\ref{tab:laces104037_phys_all}). As a result, the observed EWs should be regarded as lower limits relative to the intrinsic EWs of the young stellar component.}
    \label{fig:picket-fence Model}
\end{figure}

\subsubsection{LACES104037-LyC $f_{\mathrm{esc}}$} \label{subsubsect:LACES10_fesc}

The previous sections describe in detail the stellar population and nebular spectral modelling under the picket-fence model assumption. We constrain $f_{\mathrm{esc}}$ of LACES104037-LyC by forward-modeling the observed spectra and LyC flux, with stellar population age and $f_{\mathrm{esc}}$ being the two free parameters.

It has been extensively discussed that variations in stellar population age and $f_{\mathrm{esc}}$ occupy distinct regions in the parameter space defined by the $\beta_{UV}$ and hydrogen recombination line EWs \citep{Zackrisson_2013,Zackrisson_2017,marques-chaves_extreme_2022,Jaskot_lowz_review_2025,Giovinazzo_2025,Marques-Chaves_2026}. Figure~\ref{fig:picket-fence Model} presents the results of our model calculations, showing that different combinations of $f_{\mathrm{esc}}$ and age can be effectively distinguished in the $\beta_{UV}$--EW space. Due to the lack of high spatial resolution observations of the rest-UV SED, we only mark the ground-based measurements of LACES104037-bulk in this parameter space to constrain the youngest stellar population age of LACES104037. Given the complex and extended SFH of real galaxies, this constraint should be regarded as an upper limit.

For LACES104037-LyC, observational constraints are limited to H$\alpha$, \OIII emission lines, and the LyC band. The emission-line EWs of LACES104037-LyC are significantly lower than the average values of LACES104037-bulk. Figure~\ref{fig:picket-fence Model} illustrates the pronounced difference between LACES104037-LyC and LACES104037-bulk in the EW(H$\alpha$)--EW(\OIII) parameter space, demonstrating the impact of $f_{\mathrm{esc}}$ on emission-line EWs, namely that increasing $f_{\mathrm{esc}}$ leads to decreasing EWs.

However, as also shown in Figure~\ref{fig:picket-fence Model}, emission-line EWs alone cannot uniquely constrain $f_{\mathrm{esc}}$ due to their degeneracy with stellar age. 
The variations in EW driven by changes in age and $f_{\mathrm{esc}}$ largely follow the same evolutionary track. Therefore, if we do not consider the LyC-band emission detected in LACES104037-LyC, we would instead be inclined to interpret its extremely low EW as arising from an old stellar population with an age $ \gtrsim 1$ \,Gyr \citep{Zanella_2015}.
Ideally, under simplified assumptions (e.g., negligible dust attenuation and a cosmic mean IGM transmission), direct observations in the LyC band would allow us to constrain the escaping ionizing photon flux. Such measurements would provide the additional constraints  necessary to break this degeneracy.

\begin{figure*}[htbp]
    \centering
    \includegraphics[width=1\textwidth]{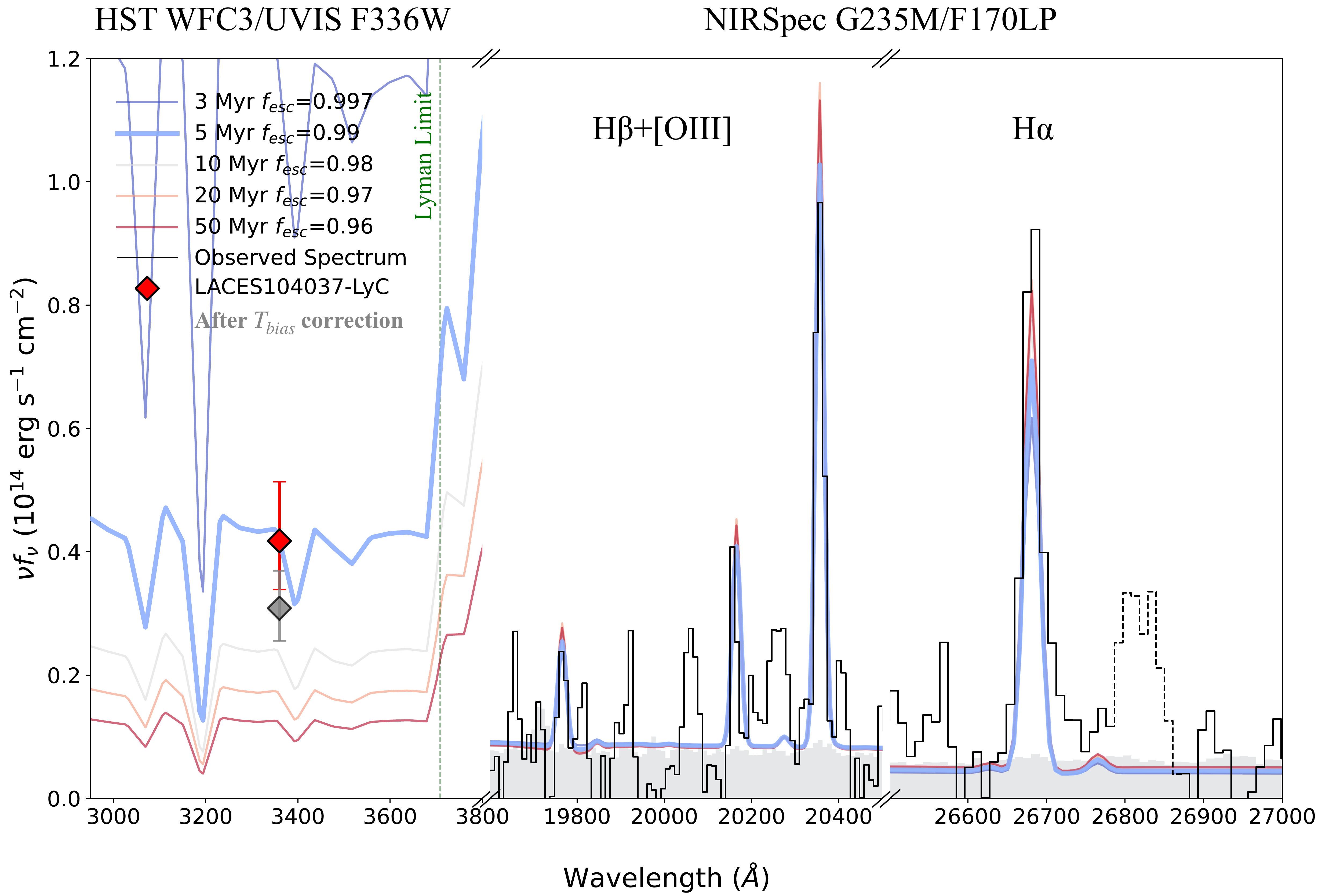}
    \caption{Forward-model fitting results for LACES104037-LyC. Observational constraints are limited to the H$\alpha$ and \OIII emission lines and the LyC band emission, obtained with sub-kiloparsec resolution observations in this LyC-leaking region. The spectral ranges affected by instrumental artifacts are indicated by dashed lines, which show substantial differences before and after the refined data reduction (see Appendix~\ref{app:JWST_data_reduction}, Figure~\ref{fig:data_processing}). We show model spectra at different stellar population ages, where $f_{\mathrm{esc}}$ is adjusted and the total flux is normalized to the integrated IFU spectrum to reproduce the observed optical spectrum. The LyC band photometry breaks the degeneracy between $f_{\mathrm{esc}}$ and stellar age present in the emission-line EW parameter space (the lower panel of Figure~\ref{fig:picket-fence Model}), and only a model with age $=5$~Myr and $f_{\rm esc}=0.99$ provides a simultaneous match to all observational constraints.}
    \label{fig:Forward-model_fitting}
\end{figure*}

We perform spectral matching of the model spectra to the observations of LACES104037-LyC. Since the model spectra represent intrinsic galaxy emission, several corrections are applied to enable comparison with observations, focusing on the wavelength range covered by the JWST IFU. First, we correct for distance dimming and cosmological redshift effects. Under the dust-free assumption for LACES104037-LyC, no internal dust attenuation is applied. Given that the SSA22 field lies toward the Galactic anti-center, the Galactic foreground extinction toward LACES104037 is $E(B-V)\sim0.05$, and its effect on the near-infrared wavelengths covered by F170LP is negligible. Within the F170LP wavelength range, we convolve the model spectra with the G235M LSF to simulate the observed line broadening. The model spectra are then normalized to the integrated luminosity in F170LP (see Figure~\ref{fig:Forward-model_fitting}).

For the LyC band, we correct the observed photometric fluxes to match the model spectra by accounting for two effects: Galactic foreground extinction and IGM absorption. Assuming the average Milky Way extinction curve \citep{Cardelli_1989}, the Galactic extinction correction factor for F336W is approximately 1.2.
For IGM absorption, we assume the cosmic mean IGM transmission at $z=3.06$. 
As discussed in \cite{Wang_Shengzhe_2025}, this correction is highly sensitive to the choice of IGM model and can introduce a factor of 2--3 uncertainty. We adopt the IGM models of \cite{inoue_updated_2014}, \cite{Steidel_2018}, and \cite{Matsuda_2005} to compute the mean IGM transmission ($\bar{t}_{\mathrm{IGM}}$) in F336W at $z=3.06$, following the methodology of \cite{wang_lyman_2025}, yielding $\bar{t}_{\mathrm{IGM}}$ values in the range of 0.179--0.215.
The corrected F336W photometry includes systematic uncertainties arising from the IGM models (see Figure~\ref{fig:Forward-model_fitting}).

We note that the use of the $\bar{t}_{\mathrm{IGM}}$ may systematically overestimate the inferred $f_{\mathrm{esc}}$ for LyC-detected galaxies, since such objects are preferentially selected along sightlines with above-average IGM transmission \citep{Bassett_2021}. Following the analysis of \citet{Bassett_2021}, we adopt an $\bar{t}_{\mathrm{IGM}}$ correction of $T_{\rm bias}=0.07$, corresponding to the F336W 5$\sigma$ detection depth of our observations. Applying this correction increases the effective transmission from $\bar{t}_{\rm IGM}=0.179$--0.215 to $\bar{t}_{\rm IGM}+T_{\rm bias}=0.249$--0.285.
The corresponding predictions are shown as gray diamond in Figure~\ref{fig:Forward-model_fitting} to illustrate the potential impact of this selection effect.

As shown in Figure~\ref{fig:Forward-model_fitting}, under the assumption of a constant SFH, by varying $f_{\mathrm{esc}}$ for different stellar population ages, we find that only a model with $f_{\rm esc}=0.99$ and a stellar population age of 5~Myr simultaneously reproduces all observational constraints. 
We note that residual discrepancies remain in matching the \OIII and H$\alpha$ emission lines, which may reflect differences between the physical conditions of LACES104037-LyC and the average properties of LACES104037-bulk, primarily manifested in the \OIII/H$\alpha$ line ratio. 
In this best-fitting model, the inferred stellar mass of LACES104037-LyC is $\sim10^{7.3}\,M_\odot$.

\section{DISCUSSION}\label{sect:discussion}

\subsection{Selection Criteria of LyC Leakers}

The search for LyC leakers at high redshift has long lacked a unified and well-defined identification criterion.
In practice, sources are often classified as high-$z$ LyC leakers if a statistically significant signal is detected in the LyC band corresponding to a given redshift, and if this signal is spatially consistent with a spectroscopically confirmed high-$z$ galaxy \citep{Fletcher_2019}.
However, such identifications are subject to substantial uncertainties. At high redshift, absorption by the IGM severely attenuates LyC photons, rendering the LyC-band emission intrinsically faint \citep{inoue_updated_2014}. 
In addition, contamination from low-redshift interlopers can lead to false-positive detections \citep{Me_tri__2025}. 
As a result, the search for and confirmation of high-$z$ LyC leakers critically depend on two key factors: (1) the depth of the LyC-band and spectroscopic observations, and (2) the angular resolution across different wavebands.

\subsubsection{Observational Depth}

The impact of LyC-band depth on the selection of LyC leaker samples has been extensively discussed in previous studies \citep{wang_lyman_2025,Rutkowski_2017}. Owing to limited observational depth, however, most of these works can only place upper limits on the $f_{\mathrm{esc}}$, rather than providing direct measurements. This limitation is significantly alleviated in the SSA22 protocluster field, which benefits from the deepest LyC-band observations available to date. The field has been observed with 20 orbits of HST/WFC3 F336W imaging, reaching a depth of 29.8\,mag ($5\sigma$ within a 0.12\arcsec\ aperture). 
Taking LACES94460 as an example, following \cite{wang_lyman_2025} method,although no LyC emission is detected, under the assumption of an average IGM transmission, the achieved depth enables a $5\sigma$ constraint on the $f_{\mathrm{esc}}$ of LACES94460 down to $f_{\mathrm{esc}} \lesssim 0.2$, making it a uniquely powerful dataset for LyC studies at high redshift.
It should be noted, however, that any selection based on LyC-band luminosity inherently introduces selection biases, preferentially identifying strong LyC leakers. This selection effect differs fundamentally from that of the LzLCS sample at low redshift \citep{Flury_2022_I}. At high redshift, current LyC-band depths generally preclude the detection of LyC leakers with low $f_{\mathrm{esc}}$, making the resulting samples intrinsically biased toward systems with extreme $f_{\mathrm{esc}}$.

Spectroscopic depth is also a crucial factor in the secure identification of LyC leakers. 
As discussed in Section~\ref{subsect:LyC_D_contamination}, LACES94460 is contaminated by a faint foreground source at $z=1.59$. 
Within the LACES program, LACES94460 has Keck/MOSFIRE $H$ and $K$ band spectroscopic coverage \citep{Nakajima_2020}. 
Although ground-based observations cannot spatially resolve the two sources, a sufficiently bright foreground interloper would, in principle, reveal itself through unassociated emission lines in the spectrum.

In the case of LACES94460, the H$\alpha$ emission line of the LACES94460-b falls within the MOSFIRE $H$ band wavelength coverage. 
However, the line flux lies below the detection limit of the Keck observations, leaving no identifiable signature. 
As a result, LACES94460 was initially misclassified as a LyC leaker, illustrating how limited spectroscopic depth and wavelength coverage can lead to false-positive LyC detections.

On the other hand, as demonstrated in Section~\ref{subsect:SFH_Model}, at constant SFH and stellar population age, increasing $f_{\mathrm{esc}}$ leads to systematically weaker nebular emission lines. 
This effect places stringent requirements on spectroscopic depth, particularly for LyC leaker studies at high redshift. 
Recently, \citet{Marques-Chaves_2026} reported a LyC leaker candidate at $z=10.255$, U37126, characterized as a very blue, ISM-naked starburst, highlighting the critical importance of spectroscopic depth. 
The key observational features of this source are its extremely steep $\beta_{UV}$ combined with very weak nebular emission. 
Even with 11 hours of JWST/MIRI low-resolution spectroscopic observations, no significant rest-frame optical emission lines are detected. 
In the absence of a prominent Lyman break, the spectroscopic redshift of such a source cannot be robustly established based solely on emission-line diagnostics.
Deep spectroscopy is therefore essential not only to identify faint foreground contaminants, but also to robustly confirm genuine LyC leakers whose nebular emission lines may be intrinsically weak due to high $f_{\mathrm{esc}}$.

\subsubsection{Angular Resolution}

The impact of angular resolution on the identification of LyC leakers has remained relatively unclear. 
One of the most critical issues is contamination from angularly close low-redshift interlopers. 
A representative example is Ion3: ground-based observations lacked sufficient spatial resolution to resolve contamination from a low-$z$ interloper within the seeing disk \citep{Me_tri__2025}. 
Only with high-resolution HST LyC-band imaging was it confirmed that Ion3 is indeed contaminated, with an overall LyC-band detection significance of only $3.5\sigma$, far below earlier expectations. 
This case highlights the importance of high spatial resolution LyC-band observations and demonstrates that LyC detections based on low-resolution data must be interpreted with caution.

On the other hand, high spatial resolution observations have revealed a substantial population of LyC leaker candidates exhibiting significant spatial offsets between the LyC emission and the stellar component, many of which show merger-like morphologies. 
For example, \citet{zhu_lyman_2024} reported that LyC leaker candidates with merger features in their sample typically exhibit offsets of $\sim$0.5\arcsec. 
Furthermore, \citet{yuan_merging_2024} and \citet{gupta_mosel_2024} identified two merger systems with high-significance LyC-band detections and clear spatial offsets, for which the probability of contamination by low-$z$ interlopers is estimated to be below 1\%. 
However, these studies lack sub-kiloparsec resolution spectroscopic observations, preventing a definitive confirmation of the LyC origin of the detected emission. 
Nevertheless, these observational findings naturally raise the question of whether merging systems can produce LyC escape regions that are spatially offset from  the main galaxy body.

LACES104037 and LACES94460 provide two instructive and complementary case studies. 
As shown in Section~\ref{subsect:LyC_D_contamination}, LACES104037 exhibits a large LyC-band offset of $\sim$0.6\arcsec ($\sim$4.6 kpc), yet IFU observations unambiguously confirm it as a merger-driven LyC leaker, with a locally inferred escape fraction reaching up to $\sim$99\%. 
This represents one of the clearest confirmations to date that merger-driven LyC leakage can arise from regions spatially offset from the stellar component, such as within a gaseous bridge connecting interacting galaxies.
Therefore, the growing population of LyC leaker candidates with spatial offsets and merger signatures is unlikely to be purely coincidental.

In contrast, LACES94460 exhibits a smaller LyC-band offset of only $\sim$0.3\arcsec ($\sim$2.3 kpc). 
Although its redshift is spectroscopically confirmed by Keck observations, it is ultimately shown to be contaminated by a faint low-$z$ interloper. 
Actually, ground-based optical observations are often insufficient to resolve the spatial separation between genuine LyC offsets and low-$z$ interlopers, unless adaptive optics are employed to mitigate seeing effects.
In imaging data, LACES94460 forms a close pair with a source at $z \sim 2.6$. 
While no prominent merger features are evident, the imaging alone does not allow us to rule out a merging scenario. 
The case of LACES94460 thus illustrates the intrinsic complexity of identifying LyC leakers: without sub-kiloparsec resolution spectroscopic confirmation, it is not possible to unambiguously classify a source as a genuine LyC leaker.

In summary, the most secure “gold standard” for identifying LyC leakers requires 
(1) a $\geq5\sigma$ detection in high–spatial–resolution LyC-band imaging, 
together with 
(2) sub-kiloparsec–resolution spectroscopic confirmation of the redshift at the same location. 
These stringent requirements inevitably make the search for LyC leakers at high-$z$ an observationally expensive endeavor.

The most efficient observational strategy therefore combines deep LyC-band imaging with wide-field slitless spectroscopy coverage, enabling blind and statistically complete sample selection. 
The MMAMOTH survey provides a prime example of such an approach \citep{Zhou_hang_2025,Wang_Shengzhe_2025,Xin_Wang_2022_MAMMOTH}. 
The MAMMOTH-grism program delivers high spatial-resolution emission-line maps, providing the high spatial-resolution redshift information required by the MAMMOTH-LyC program. 
Through this combined observational strategy, systems like LACES94460 that are contaminated by low-$z$ interlopers can be reliably distinguished from genuine LyC leakers.

Integral-field spectroscopy, on the other hand, is best suited for follow-up observations of LyC leaker candidates selected from imaging surveys, particularly for systems exhibiting spatial offsets. 
IFU observations allow both secure redshift confirmation and detailed kinematic analyses, which are essential for understanding the physical origin of LyC escape.
The contrasting cases of LACES104037 and LACES94460 clearly illustrate the necessity of this two-step strategy and highlight the risks of LyC leaker identification based on incomplete observational constraints.

\subsection{Star Formation Model} \label{subsect:SFH_Model}

A key effect of ionizing photon escape is the suppression of nebular emission relative to the stellar continuum, particularly in systems with high  $f_{\mathrm{esc}}$ \citep{Giovinazzo_2025,Marques-Chaves_2026}.
In the absence of ionizing photon escape, the strength of nebular emission relative to the stellar continuum is strongly correlated with stellar population age \citep{Zanella_2015}. 
However, once ionizing photons escape, the energy available to power nebular emission is reduced, leading to weaker nebular radiation for a given stellar population.
In observations outside the LyC band, this effect primarily manifests in two ways: (1) weakened nebular emission lines, and (2) a reduced reddening contribution of the nebular continuum relative to the stellar spectrum \citep{marques-chaves_extreme_2022,Narayanan_2025}. 
We note that several studies have shown that the relationship between emission-line EW and $f_{\mathrm{esc}}$ in LyC leakers is not necessarily anti-correlated \citep{Flury_2022_I,Izotov_2016}, and strong nebular emission lines can still be observed in LyC-emitting systems \citep{rivera-thorsen_gravitational_2019,Naidu_2017}. However, such statistical samples are often affected by variations in SFH and anisotropies in the ISM geometry \citep{Alavi_2020,Flury_2022_I}, which can complicate the interpretation of the observed trends. Nevertheless, for systems approaching very high $f_{\mathrm{esc}}$, nebular suppression becomes increasingly significant due to the reduced nebular energy budget \citep{Zackrisson_2013}.
This energy-budget effect underlies the separation induced by $f_{\mathrm{esc}}$ in the parameter space defined by emission-line EW and the $\beta_{UV}$ (Figure~\ref{fig:picket-fence Model}). 

To illustrate this behavior, we adopt the picket-fence model as a simplified description of LyC escape. The primary motivation for this choice is that the current observations provide no direct constraints on the neutral hydrogen column density ($N_{\rm H}$) along the LyC escape channels. More sophisticated density-bounded models would require exploring a wide range of $N_{\rm H}$ values, together with additional assumptions regarding the gas geometry and density structure. Because these quantities cannot be independently constrained by the available data, such models remain highly degenerate and would introduce substantially more free parameters without providing correspondingly stronger observational constraints.
The picket-fence model should therefore be regarded as a first-order approximation rather than a physically complete description of the escape process. Its advantage is that it captures the dominant observational consequences of LyC leakage while remaining sufficiently simple to be constrained by the existing data. This framework has consequently been widely adopted in studies of LyC leakers (e.g., \citealt{izotov_detection_2016}; \citealt{marques-chaves_extreme_2022}; \citealt{Marques-Chaves_2026}). We emphasize that the true escape geometry is likely to be considerably more complex than represented by this idealized model. Future observations capable of directly constraining the $N_{\rm H}$ or resolving the distribution of neutral gas along the escape channels will be essential for distinguishing between different escape scenarios and for developing a more physically realistic understanding of ionizing photon escape in LACES104037.

To directly constrain $f_{\mathrm{esc}}$ using the picket-fence model, a specific SFH must be assumed. 
Since the recent SFH of very young stellar populations is generally difficult to constrain observationally, we adopt a constant SFH as a simple and commonly used parameterization, consistent with previous studies (e.g., \citealt{izotov_detection_2016,Giovinazzo_2025,Marques-Chaves_2026}). 
As discussed in \cite{izotov_detection_2016}, under a constant SFH, the LyC photon production rate reaches an approximately steady state after $\sim5$~Myr, and the resulting nebular emission remains nearly unchanged. However, as stellar age increases, the accumulation of low-mass stars leads to a continuous rise in the optical continuum intensity for a fixed $f_{\mathrm{esc}}$. This evolution primarily affects the EWs of optical emission lines. Consequently, the relative strength between the LyC band and the optical continuum increases with stellar age. Therefore, once both the optical emission-line EWs and the LyC band signal are observed, stellar age and $f_{\mathrm{esc}}$ can be uniquely determined under the assumptions of the picket-fence model and a constant SFH.
Different SFH prescriptions may shift the inferred age and escape fraction, representing an additional source of systematic uncertainty.

It should be further noted that the models adopted in this work do not account for the contribution of an underlying old stellar population. In real galaxies, star formation is expected to have occurred prior to the merger event. Although such early star formation may not have been particularly intense, the long-term accumulation of low-mass stars can significantly enhance the optical continuum and consequently reduce the observed emission-line EWs. 
As a result, the emission-line EWs measured for LACES104037-bulk may be systematically lower than the values predicted by models that assume a purely young stellar population. Therefore, the discrepancy between the observed EWs and model predictions may be partly driven by continuum dilution from an underlying older stellar population (see Figure~\ref{fig:picket-fence Model}).

For LACES104037-LyC, however, this effect is likely less severe than for LACES104037-bulk. Studies in the local universe suggest that star formation in tidal tails typically occurs over relatively short timescales \citep{Mulia_2015,Neff_2005}, which limits the buildup of a substantial low-mass stellar population. In some systems, such as NGC~3447, the tidal structures are found to host little to no old stellar component \citep{Riess_2025}. We therefore argue that neglecting a potential old stellar population in LACES104037-LyC is a reasonable approximation and is unlikely to introduce significant changes to our results.

At high redshift, an unavoidable challenge in determining $f_{\mathrm{esc}}$ is IGM absorption. As discussed in \cite{Wang_Shengzhe_2025}, for an individual LyC leaker, this effect is intrinsically degenerate with $f_{\mathrm{esc}}$. In principle, the IGM transmission along a specific line of sight cannot be directly determined. Instead, one typically relies on IGM models to estimate a cosmic mean transmission, which is then used to correct the observed LyC band photometry. In Section~\ref{subsubsect:LACES10_fesc}, we adopt three different IGM models and incorporate the systematic differences among their correction factors into the overall uncertainty. It is important to emphasize that this procedure accounts for the uncertainty arising from different IGM models in predicting the cosmic mean transmission, rather than the variation of IGM absorption along the specific line of sight toward LACES104037. In reality, the IGM absorption toward LACES104037 could be either stronger or weaker than the cosmic mean. 
In addition, LyC-detected galaxies are not expected to represent a random subset of all sightlines. Because LyC emission is more readily detected along relatively transparent sightlines, the average IGM transmission for detected sources may be systematically higher than the cosmic mean. Following \citet{Bassett_2021}, we estimate a transmission bias of $T_{\rm bias}=0.07$ for the F336W 5$\sigma$ detection depth of our observations. This effect is illustrated in Figure~\ref{fig:Forward-model_fitting} (gray diamond) and would shift the inferred transmission toward higher values than the cosmic mean. Nevertheless, the magnitude of this correction remains modest compared to the intrinsic sightline-to-sightline variation in IGM absorption.
Stronger IGM absorption would favor higher $f_{\mathrm{esc}}$ and younger stellar ages, while weaker absorption would favor lower $f_{\mathrm{esc}}$ and older stellar ages. In this sense, our method does not fully break the degeneracy between IGM absorption and $f_{\mathrm{esc}}$ for LACES104037-LyC. Nevertheless, we emphasize that with additional observational constraints on the SFH and stellar age, it is in principle possible to alleviate this degeneracy between IGM transmission and $f_{\mathrm{esc}}$ within the framework of the picket-fence model.

For the EoR and even higher redshifts, diagnostics of $f_{\mathrm{esc}}$ in the parameter space defined by the $\beta_{UV}$ and optical emission-line EWs hold significant promise. 
A recent and illustrative example is U37126, for which $f_{\mathrm{esc}}$ can be inferred based on these observables \citep{Marques-Chaves_2026}.
However, such inferences are subject to substantial limitations, as they depend sensitively on assumptions about dust attenuation and the star formation history. 
For metal emission lines, the diagnostics are further affected by additional physical parameters, most notably metallicity. 
As a result, for LyC leakers with relatively low $f_{\mathrm{esc}}$, this approach can suffer from large systematic uncertainties. 
Nevertheless, we argue that this method remains highly effective for identifying strong LyC leakers.

\subsection{Merger and LyC Escape}\label{subsect:merger}

Whether merger systems play a significant role in facilitating ionizing photon escape has long been a subject of debate (e.g., \citealt{zhu_lyman_2024,yuan_merging_2024,Weilbacher_2018}). Numerous studies have suggested that the merger process can have a substantial impact on LyC escape. In particular, at high redshift, a large number of spatially offset LyC band detections have been reported in merger systems \citep{zhu_lyman_2024,gupta_mosel_2024,yuan_merging_2024}. Although a non-negligible fraction of these detections may be contaminated by low-$z$ interlopers, these studies collectively raise an important question: whether star formation activity induced by galaxy interactions and occurring outside the main body of galaxies, beyond the half-light radius, tends to exhibit systematically higher $f_{\mathrm{esc}}$.

In the local universe, the lack of instrumentation capable of directly detecting emission in the LyC band has severely limited observational constraints \citep{Jaskot_lowz_review_2025}. Although several studies have carefully discussed the possibility of ionizing photon escape from tidal structures in the nearby ``Antennae Galaxy'' (NGC~4038/39; \cite{Weilbacher_2018}), to date there is still no direct observational evidence in the local universe confirming the presence of LyC escape from such regions.

The importance of LACES104037 lies in the fact that it provides direct evidence that, at high redshift, merger-driven star formation occurring in external galactic substructures can produce extremely high ionizing photon escape fractions, with $f_{\mathrm{esc}}$ approaching unity. To date, this remains the only confirmed case of such extreme LyC escape at high redshift. 
Interestingly, LACES104037-bulk also exhibits very recent and intense star formation within the interacting system; however, it shows no clear evidence for LyC photon escape. As shown in Figure~\ref{fig:picket-fence Model}, its dust-corrected $\beta_{\rm UV}$ suggests that a substantial $f_{\mathrm{esc}}$ would be expected in the absence of dust attenuation. This contrast with the LACES104037-LyC region implies that interaction-induced substructures may provide more favorable conditions for ionizing photon escape than the main body of a galaxy, potentially owing to a more dust-free environment.

Another representative example is J1244-LyC1, a well-studied LyC leaker at high redshift \citep{Wang_Shengzhe_2025}. J1244-LyC1 is a major merger system with a total stellar mass of $\sim10^{10.2}\,M_{\odot}$. Its star formation activity lies slightly above the star-forming main sequence, yet it exhibits prominent LyC escape with multiple spatially distinct escape regions associated with different galactic components. LACES104037 and J1244-LyC1 therefore appear to represent two distinct merger-driven LyC escape mechanisms. LACES104037 has a stellar mass within the typical range of known LyC leakers ($10^{8}$--$10^{9.5}\,M_{\odot}$; \cite{Flury_2022_I}), and is observed in an early-stage merger. In contrast, J1244-LyC1 is a massive galaxy beyond the canonical LyC leaker mass range and is observed in a late-stage merger.

Together, these two systems demonstrate that extremely efficient LyC escape can occur throughout the entire merger sequence. During the early stages, galaxy interactions can strip star-forming gas from the main bodies of galaxies, leading to star formation in substructures such as tidal features that dominate the ionizing photon escape. In the late stages of mergers, violent gas exchange between the interacting galaxies not only triggers intense star formation but also violently reshapes the ISM, enabling LyC escape through multiple channels and across multiple structural components.

\subsection{Merger-driven Reionization}

A key and particularly intriguing question concerns the connection between the EoR and merger-driven LyC leakers.
Over the past two decades, numerous studies have attempted to constrain the ionization history of the EoR  (e.g., \citealt{Mascia_2025,Austin_2025,Wu_2025,Jecmen_2026,Duncan_2015}), largely within the framework of the reionization models originally proposed by \citet{Madau_1999}.
These models are primarily built upon the galaxy UV luminosity function, with the $f_{\mathrm{esc}}$ representing the dominant source of uncertainty.
In most existing studies, $f_{\mathrm{esc}}$ is either assumed to be constant or indirectly constrained using empirical relations derived from the low-redshift LyC leaker sample (e.g., LzLCS;\cite{Flury_2022_I}).

However, such an approach implicitly assumes that the physical mechanisms governing LyC escape are uniform across the galaxy population and operate in isolation, for example through feedback-driven channels such as supernovae \citep{Jaskot_lowz_review_2025, carr_2025}.
The indirect relations established by LzLCS are likewise calibrated under this assumption.
This simplification neglects the intrinsic diversity of LyC escape mechanisms.
As discussed in Section~\ref{subsect:merger}, LyC escape depends not only on the ionizing photon production rate, but also critically on the surrounding ISM conditions \citep{Ji_dust_2025}.
Galaxy interactions represent one of the most effective processes capable of globally reshaping the ISM on galactic scales \citep{garaysolis_2025,puskás_2025,Purkayastha_2022_GPs}.


However, to date, only three merger-driven LyC leakers---Haro~11, LACES104037, and J1244-LyC1---have been observed with sub-kiloparsec spatial resolution \citep{Wang_Shengzhe_2025,Komarova_2024}. This limited sample size is insufficient to establish a statistically robust characterization of LyC escape mechanisms associated with galaxy mergers.
Despite the small number, these systems share a common feature: the regions exhibiting LyC leakage are spatially offset from the main stellar bodies of their host galaxies. This suggests that LyC escape in merger-driven systems may be governed primarily by localized star-forming activity and the surrounding ISM conditions, rather than by the global properties of the merging galaxies themselves.
Such spatially localized escape channels are often overlooked in prevailing analytical frameworks, which typically assume that LyC escape is a galaxy-wide phenomenon \citep{Mascia_2025,Austin_2025,Wu_2025,Jecmen_2026}. Although \citet{Mascia_2025} placed constraints on the contribution of mergers to reionization and concluded that mergers are unlikely to play a dominant role during the EoR, we argue that merger-driven LyC leakers require a more refined treatment that explicitly accounts for their spatially inhomogeneous and transient nature.

In order to incorporate merger systems into a physically motivated EoR framework, several key questions must be addressed:
(1) What fraction of LyC leakers are associated with galaxy mergers?
(2) How does the merger rate evolve with redshift?
(3) What range of $f_{\mathrm{esc}}$ is produced by merger systems, and which observables can reliably trace it?
(4) Do all merger systems exhibit LyC escape, and over what timescales does leakage persist?
(5) How do mergers affect the total ionizing photon production?

For the first question, substantial uncertainties remain. The LaCOS survey \citep{Le_Reste_2025_III} reported that more than 41\% of confirmed LyC leakers exhibit morphological features indicative of galaxy mergers. However, this fraction is likely affected by strong selection biases, as the LzLCS sample is preferentially selected to resemble Green Pea-like systems \citep{Flury_2022_I}. 
At higher redshift, \citet{zhu_lyman_2024} compiled a sample of LyC-leaking candidates, the majority of which show merger-like features. Nevertheless, these candidates generally suffer from low S/N and are subject to potential contamination by low-redshift interlopers, rendering the inferred merger fraction uncertain. 
Despite these limitations, both studies consistently suggest that galaxy mergers constitute a non-negligible fraction of the LyC-leaking population.

The second question is the most tractable observationally and is a primary science driver of JWST.
Recent studies (e.g., \citealt{Duan_2025,Calabr_2026}) indicate a high $\mathcal{R}_{\mathrm{M}}$ during the EoR, with the $\mathcal{R}_{\mathrm{M}}$ being nearly an order of magnitude higher than that at cosmic noon.
As JWST observations continue to expand in both depth and area, increasingly robust constraints on the merger rate at $z \gtrsim 6$ are expected.

At present, the remaining questions cannot be robustly addressed observationally.
Although merger-driven LyC leakers have now been identified and extreme local escape fractions have been measured, the current sample size is insufficient to establish statistically meaningful constraints on their occurrence, efficiency, or duty cycle.
We therefore emphasize the urgent need to build a dedicated sample of merger-driven LyC leakers, complemented by high-resolution simulations capable of constraining their physical mechanisms and leakage timescales.

We conclude that merger-driven reionization is a plausible scenario, but one that remains poorly constrained by existing observations.
Compared to LyC escape in isolated galaxies, merger-driven leakage may represent a more efficient and physically natural pathway.
This channel has long been overlooked, but the discoveries of LACES104037--LyC and J1244--LyC1 demonstrate that it warrants serious consideration in future models of cosmic reionization.

\subsection{Comparison with \cite{Rivera-Thorsen_2025}}
Prior to this work, \cite{Rivera-Thorsen_2025} reported the JWST observations of LACES104037. 
Compared to their study, the present work provides several key improvements and extensions: 
(1) a refined data reduction procedure that carefully addresses contaminations not fully treated by the standard JWST pipeline (Appendix~\ref{app:JWST_data_reduction}); 
(2) a robust astrometric calibration ensuring accurate spatial alignment among the HST/F336W, HST/F160W, and JWST/NIRSpec IFU data; 
(3) a confirmation that LACES94460 is contaminated by a low-redshift interloper; 
(4) a detailed determination of the physical properties of LACES104037 based on spatially resolved IFU measurements; and 
(5) a comprehensive stellar population modeling of LACES104037-LyC, leading to a more reliable constraint on its stellar age and $f_{\mathrm{esc}}$.
Below, we focus on two major differences between this work and \cite{Rivera-Thorsen_2025}: the data reduction strategy and the measurement of $f_{\mathrm{esc}}$.

At the data reduction level, we emphasize that careful treatment of instrumental effects is essential for deriving reliable results for LACES104037-LyC.
As demonstrated in Appendix~\ref{app:JWST_data_reduction} and Figure~\ref{fig:data_processing}, instrumental artifacts dominate the continuum emission beyond 2.7~$\mu$m in the original data products, which would severely bias the measurement of the optical continuum.
In addition, an anomalous emission feature near \NII~$\lambda6583$ and H$\beta$ is identified as contamination from a ''snowball'' event that was not fully masked by the standard JWST pipeline.
After dedicated cleaning, the affected spectral regions recover physically reasonable behavior.
If left uncorrected, such contamination could directly impact both the redshift identification and the inferred physical parameters of LACES104037-LyC.

Accurate astrometry is another critical prerequisite for confirming a LyC leaker.
The IFU observations of both LACES104037 and LACES94460 lack contemporaneous astrometric calibration exposures, requiring particular care in post-processing.
We think that performing astrometric alignment using only the F336W image and the IFU data is not sufficiently robust.
Without an intermediate-band constraint, it is difficult to unambiguously identify which substructure is responsible for the LyC emission, introducing substantial subjectivity.
By jointly incorporating the HST/F160W image, we achieve a more reliable astrometric solution with well-controlled uncertainties (see Appendix~\ref{app:Astrometry}).
We note that the H$\alpha$ EW of LACES104037-LyC reported by \cite{Rivera-Thorsen_2025} is systematically higher than our measurement, for which we derive an H$\alpha$ EW of $88^{+63}_{-27}\,$\AA.
This discrepancy may arise from several factors, including small astrometric offsets that lead to differences in aperture position, variations in the treatment of potential over-subtraction during the master background subtraction, and differences in the treatment of spurious line-like features, among other factors.

Finally, proper treatment of observational uncertainties is essential.
As shown in Figure~\ref{fig:LACES104037_data}, even after applying frame-by-frame error corrections to the IFU data, the extracted spectral uncertainties remain a factor of $\sim$2--3 larger than the {\tt ERR} cube noise level.
We explicitly account for this effect in our emission-line measurements and subsequent analysis.
Although this leads to lower formal S/N, it ensures that our line measurements are statistically reliable.

Regarding the measurement of $f_{\mathrm{esc}}$, we find that the estimate presented in \cite{Rivera-Thorsen_2025} is subject to systematic uncertainties.
By definition, $f_{\mathrm{esc}}$ depends on the intrinsic LyC spectral shape, and the conversion from LyC-band imaging to the intrinsic ionizing photon production rate $Q(\mathrm{LyC})$ varies significantly with the assumed stellar population model \citep{Jaskot_lowz_review_2025}.
Consequently, the choice of stellar population model can significantly affect the inferred value of $f_{\mathrm{esc}}$.
In addition, the inferred $f_{\mathrm{esc}}$ depends sensitively on the assumed IGM transmission \citep{Wang_Shengzhe_2025}.
We emphasize that although our estimates of $f_{\mathrm{esc}}$, stellar population age, and stellar population properties rely on specific modeling assumptions, they are internally self-consistent and capable of simultaneously reproducing all available observational constraints for LACES104037-LyC.


\section{Summary} \label{sect:Summary }

This work presents a detailed analysis of two LyC leaker candidates selected from the LACES project in the SSA22 protocluster field at $z\sim3.06$, based on the JWST NIRSpec IFU observations. Through careful processing of the JWST IFU data combined with HST F336W (LyC band) imaging, we confirm LACES104037 as a merger-driven LyC leaker, while demonstrating that the apparent LyC signal associated with LACES94460 is contaminated by a low-$z$ interloper.

The JWST NIRSpec IFU observations establish LACES104037 as an early-stage merger system, consisting of two primary interacting galaxies, LACES104037 and LACES104037s, separated by $\sim12$~kpc with a projected velocity offset of $\sim480~\mathrm{km\,s^{-1}}$. The ionizing photon emission originates from LACES104037-LyC, which is located on a tidal-tail structure between the two merging galaxies. The JWST IFU data cover the key rest-frame optical wavelength range ($\sim4200$--$7800$~\AA), providing a rich set of emission-line diagnostics, including \OIII~$\lambda4363$ and the \SII doublet. Combined with Keck $H$ band observations, we perform robust plasma diagnostics for the entire LACES104037 system, deriving its average gas-phase metallicity, dust attenuation, and other physical parameters. In addition, incorporating ground-based Subaru observations, we carry out SED fitting to constrain the global physical properties of LACES104037.

The observational properties of LACES104037-LyC differ markedly from those of LACES104037-bulk, with emission-line EWs that are significantly lower than the galaxy-wide average. To estimate the local $f_{\mathrm{esc}}$ of LACES104037-LyC, we construct a toy model based on \texttt{Starburst99} and \texttt{MAPPINGS}, adopting the picket-fence model framework. By comparing the model predictions with both the LyC band photometry and the IFU spectroscopy, we constrain $f_{\mathrm{esc}}$ and stellar population age. Our results indicate that LACES104037-LyC is a star-forming clump located on a tidal tail, with a stellar age of $\sim5$~Myr and an extreme escape fraction of $f_{\rm esc}\sim0.99$.

LACES104037-LyC represents the first spectroscopically confirmed LyC leaker at high redshift that originates from an external star-forming clump outside the main body of its host galaxy. This discovery significantly expands our understanding of merger-driven ionizing photon escape mechanisms. Together with J1244-LyC1, LACES104037-LyC helps to establish a coherent physical picture of merger-driven LyC escape operating across different merger stages.

Despite the identification of LACES104037 and J1244-LyC1, our understanding of these systems remains far from complete. Our analysis increasingly relies on observations with high spatial resolution, as resolving individual star-forming clumps is essential for revealing the underlying physics of ionizing photon escape. This places stringent demands on observational data. On the one hand, extensive multi-band imaging and spectroscopic observations of LyC leakers with space-based facilities are required; on the other hand, targeted LyC band observations of low-redshift merger systems should be actively pursued.

More broadly, as our understanding of ionizing photon escape in galaxies continues to improve, traditional approaches that estimate the ionizing budget based solely on the UV luminosity function appear increasingly inadequate. Such methods fail to capture the complexity of processes such as merger-driven LyC escape and require more refined treatments of $f_{\mathrm{esc}}$. With the growing body of JWST observations probing the EoR, this issue is expected to be progressively resolved. 

\begin{acknowledgments}
We thank the anonymous referee for very constructive comments that helped improve the quality of this Letter.
XW thanks Christian Hayes and Brian Hilbert for their help with NIRSpec/IFU data reduction.
This work is supported by the National Key R\&D Program of China No.2025YFF0510603, the National Natural Science Foundation of China (grant 12373009), the CAS Project for Young Scientists in Basic Research Grant No. YSBR-062, the China Manned Space Program with grant no. CMS-CSST-2025-A06, and the Fundamental Research Funds for the Central Universities. XW acknowledges the support by the Xiaomi Young Talents Program, and the work carried out, in part, at the Swinburne University of Technology, sponsored by the ACAMAR visiting fellowship. AKI is supported by JSPS KAKENHI Grant No. 23H00131. TN thanks support through ARC Discovery Project Grant DP230103161.

This work is based on observations made with the NASA/ESA/CSA James Webb Space Telescope. The data were obtained from the Mikulski Archive for Space Telescopes at the Space Telescope Science Institute, which is operated by the Association of Universities for Research in Astronomy, Inc., under NASA contract NAS 5-03127 for JWST. These observations are associated with program 1827, 1869.

Some of the data presented in this paper were obtained from the Mikulski Archive for Space Telescopes (MAST) at the Space Telescope Science Institute. The specific observations analyzed can be accessed via \dataset[https://doi.org/10.17909/2hpc-w143]{https://doi.org/10.17909/2hpc-w143}. STScI is operated by the Association of Universities for Research in Astronomy, Inc., under NASA contract NAS5–26555. Support to MAST for these data is provided by the NASA Office of Space Science via grant NAG5–7584 and by other grants and contracts.

\end{acknowledgments}
%
%













\bibliography{LyC_leaker_observation}
\appendix \label{sect:app}

\section{JWST DATA PROCESSING DETAILS}\label{app:JWST_data_reduction}

Here we provide a concise overview of the overall processing steps applied to the observational data. 
The raw data were reduced using version 1.19.1 of the STScI pipeline\footnote{\href{https://github.com/spacetelescope/jwst-pipeline-notebooks/blob/main/notebooks/NIRSPEC/IFU/JWPipeNB-NIRSpec-IFU.ipynb}{JWST pipeline}} and the CRDS context \texttt{jwst\_1413.pmap}.

In the data reduction process, several non-standard settings were adopted. 
To correct for the 1/f noise, during stage 1 we applied the 
\texttt{Clean Flicker Noise}\footnote{\href{https://jwst-pipeline.readthedocs.io/en/stable/jwst/clean_flicker_noise/index.html}{JWST pipeline clean flicker noise information}}, 
and during stage 2 we enabled the 
\texttt{NSClean 1/f Correction}\footnote{\href{https://jwst-pipeline.readthedocs.io/en/stable/jwst/nsclean/index.html}{JWST pipeline NSClean 1/f Correction information}}. 
Both procedures essentially implement a 1/f noise removal algorithm \citep{rauscher2023nscleanalgorithmremovingcorrelated}. 
In stage 1, the Clean Flicker Noise is applied on a group-by-group basis before calculating the count rate image, 
while in stage 2, the NSClean 1/f Correction is applied on the rate image. 
These two corrections are not in conflict, but applying them simultaneously may lead to a certain level of flux loss. 
To ensure that the final data cube is sufficiently clean, we enabled both corrections. 

To mitigate the impact of bad pixels, we further applied the 
\texttt{Bad Pixel Self-Calibration}\footnote{\href{https://jwst-pipeline.readthedocs.io/en/stable/jwst/badpix_selfcal/index.html}{JWST pipeline Bad Pixel information}} in stage 2. 
In addition, the presence of failed open shutters during the observation produces uneven flux enhancements in the data cube. 
To eliminate this effect, after stage2 we masked the pixels flagged in the \texttt{\_cal.fits} files with DQ value 
"MSA\_FAILED\_OPEN"\footnote{\href{https://jwst-pipeline.readthedocs.io/en/stable/jwst/msaflagopen/index.html}{JWST Failed MSA information}}. 
This affects about 2\% of the data. 
Since our target is located near the center of the cube, we subsequently applied the 
\texttt{Master Background Subtraction}\footnote{\href{https://jwst-pipeline.readthedocs.io/en/stable/jwst/master_background/index.html}{JWST pipeline master background information}} 
in stage 3 to achieve a more accurate background removal. 

\begin{figure*}[htbp]
    \centering
    \includegraphics[width=1\textwidth]{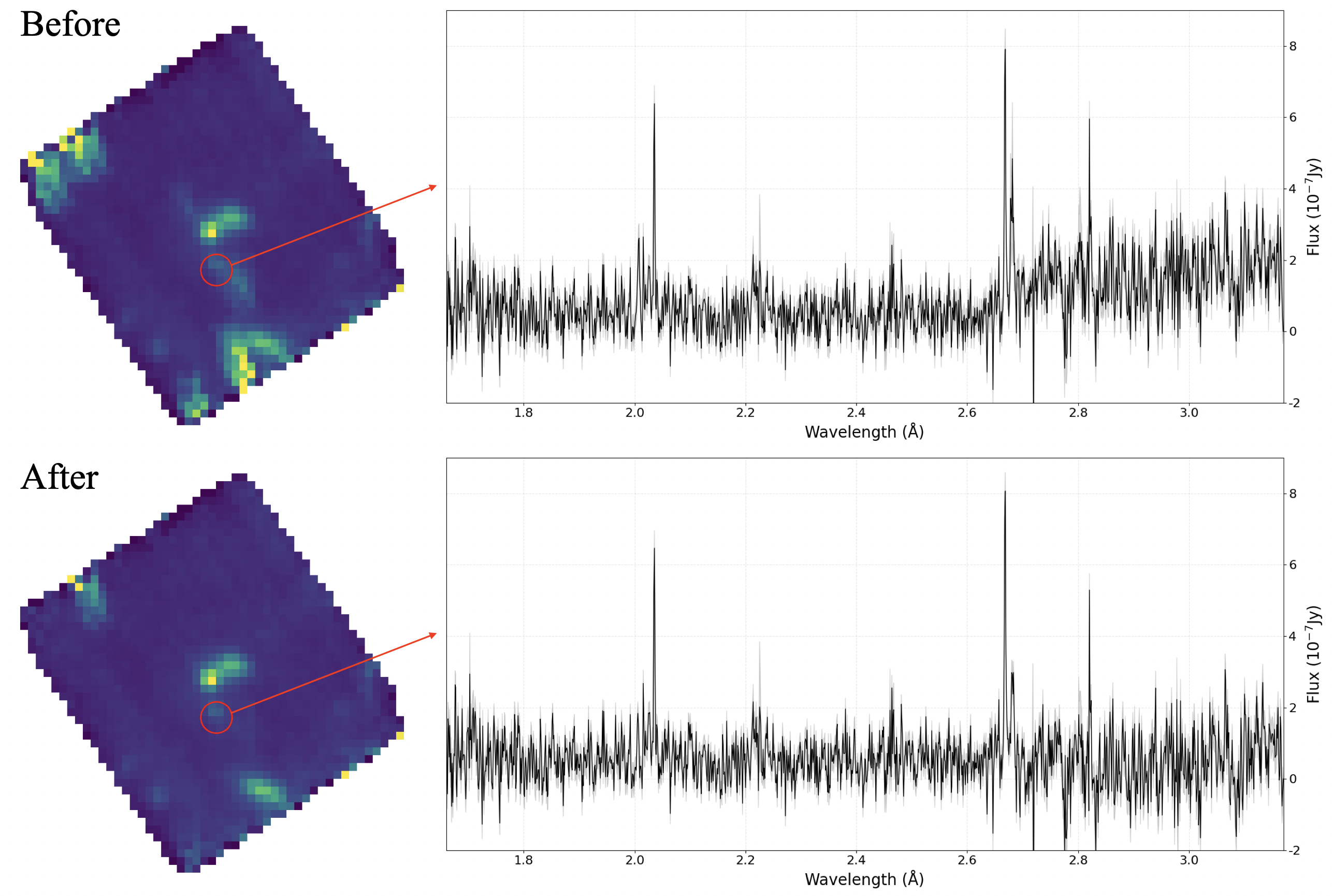}
    \caption{We compare the data quality before and after the refined data processing. Both data products were generated using master background subtraction. The refined processing introduces three major improvements. (1) The anomalous rise of the continuum beyond $2.7\,\mu$m is removed. This feature was caused by an ``MSA FAILED OPEN'' event, which predominantly affected the emission region associated with LACES104037-LyC. (2) Spurious emission-line features are eliminated. Because the JWST pipeline algorithm for removing ``snowball'' artifacts cannot fully mitigate contamination from faint residual features, we visually inspected the spectra and identified the contaminants that most strongly affect the emission lines of LACES104037-LyC. The most impacted regions are those near [NII]~$\lambda6583$ and H$\beta$. (3) Contamination at the edges of the data cube is effectively removed. The left panel shows the result obtained after stacking the optical continuum. Compared to the pre-processed data, most of the edge-related contamination is successfully mitigated. We note that, in the optical continuum image presented in the main text, the contaminant in the upper-left corner is masked and noise-filled.}
    \label{fig:data_processing}
\end{figure*}

Despite the carefully tuned pipeline processing described above, there remain unrecognized contaminants, such as incompletely removed snowballs. 
To address this, we visually inspected the secondary data products and manually masked the affected regions, ensuring that these contaminants do not impact our scientific objectives (see Figure~\ref{fig:data_processing}). 

As discussed in \citet{Fujimoto_2025}, we similarly find that the \texttt{ERR} extension of the Level~3 data products requires rescaling, because the 1$\sigma$ noise level estimated from the background regions in each science cube frame is inconsistent with the median value of the corresponding frame in the pipeline-generated error cube. In particular, the noise values provided in the \texttt{ERR} cube appear to be systematically underestimated.
Following the procedure described in \citet{Fujimoto_2025}, we apply a correction to the error arrays, with rescaling factors ranging from 1.2 to 1.3.
In addition, we find that the master background subtraction slightly over-subtracts the background. 
We apply a median background correction to each individual frame, ensuring the reliability of the extracted spectral information.

\section{NIRSpec IFU Astrometry} \label{app:Astrometry}

Accurate astrometry is a prerequisite for studying LyC leakers. 
In the observation design of GO 1827, the WATA mode was not used for auxiliary position calibration. 
Considering the widely existing issue of observational offsets in the MSA, 
we must introduce external data from the same field with already accurate astrometry for auxiliary calibration. 

For LACES94460, since JWST NIRCam F277W imaging data are available, and the wavelength coverage of F277W overlaps with that of G235M/F170LP. 
We convolved the transmission curve of F277W with the IFU data cube of LACES94460, 
thus generating an image that can be used for astrometric calibration. 
After removing spurious sources caused by instrumental contamination, 
we completed the astrometric alignment. 

For LACES104037, however, the situation is relatively more complex. 
Due to the lack of high-resolution spatial observations with effective band overlap, 
we could not perform astrometric calibration in the same way as for LACES94460. 
Instead, we masked the emission lines and contamination in the IFU data cube of LACES104037, 
and stacked the remaining continuum , 
thus generating a rest-frame 4100--7500\,\AA\ continuum image. 
We assumed that the HST F160W image of LACES104037 
(rest-frame $\sim 3900$\,\AA\ at $z \sim 3.1$) is dominated by the continuum emission from HII regions. 
Therefore, in principle, the spatial positions in the two images should be consistent. 
Based on this, we completed the astrometric calibration for LACES104037. 

\section{SED Results} \label{app:SED_Results}

\begin{figure*}[htbp]
    \centering
    \includegraphics[width=1\textwidth]{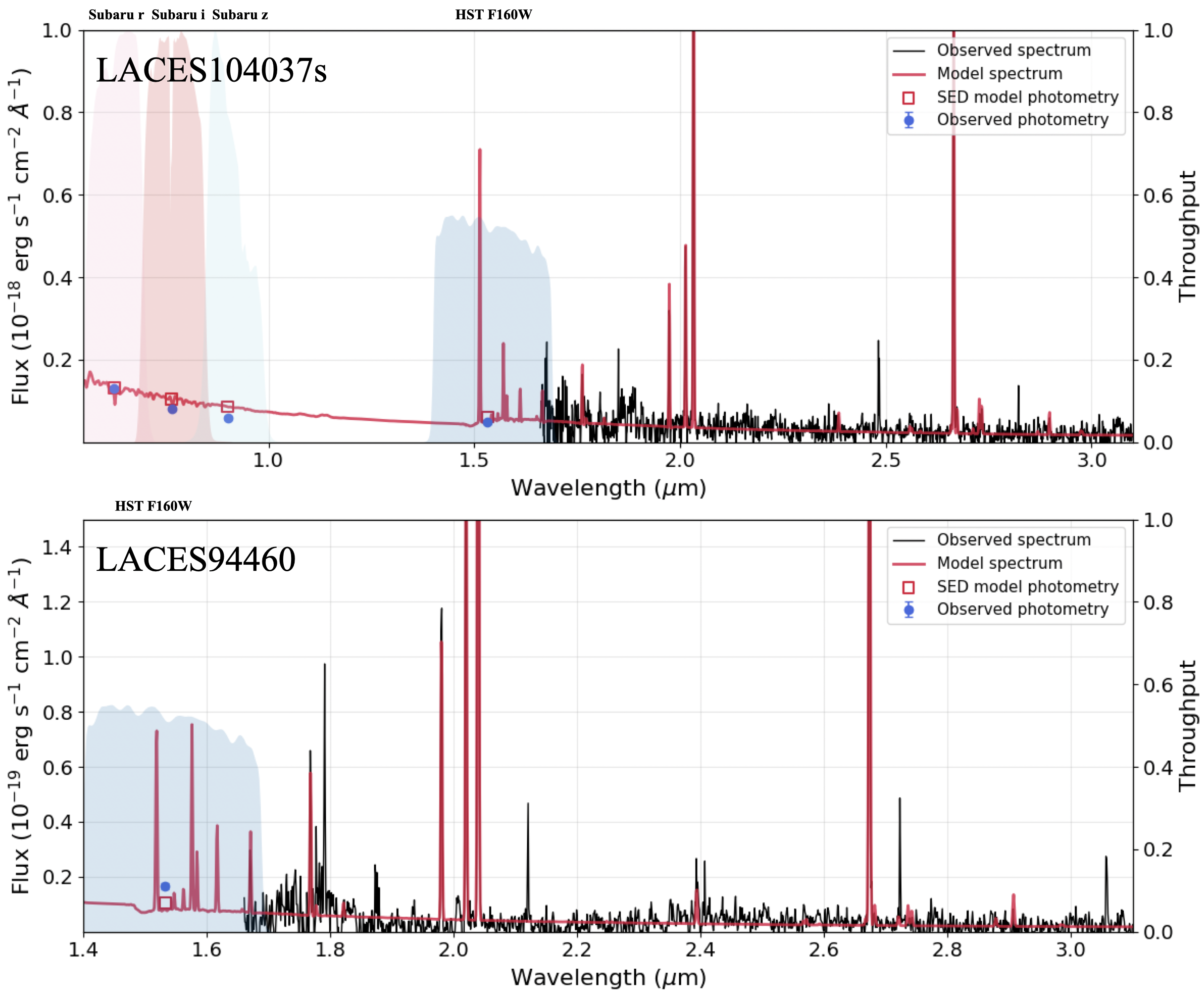}
    \caption{Best-fit SED model of LACES104037s and LACES94460 derived using \texttt{PROSPECTOR}, fitted to the existing broad-band photometry and IFU spectroscopy. Our spectral fitting results are shown in Table~\ref{tab:laces104037s_and_laces9_phys_all}. 
    } 
    \label{fig:SED_result2}
\end{figure*}

Here we present the SED fitting results for LACES104037s and LACES94460 (see Figure~\ref{fig:SED_result2} and Table~\ref{tab:laces104037s_and_laces9_phys_all}).
For LACES104037s, the source is located near the edge of the IFU FoV, and therefore the fitting may be affected by instrumental edge effects. In addition, a small fraction of the source falls outside the FoV. We correct for this missing flux using the spatial light distribution traced by the F160W image.
LACES94460, on the other hand, is an extremely faint source whose spectrum exhibits several faint emission features. Most of these features are attributable to instrumental artifacts rather than genuine emission lines. Nevertheless, these spurious features do not significantly affect the resulting SED fitting.

\begin{deluxetable}{lccccc}
\tablecaption{Broad-band photometry of the targets}  \label{tab:photometry}
\tablehead{
\colhead{Object} &
\colhead{$V$} &
\colhead{$R$} &
\colhead{$i'$} &
\colhead{$z'$} &
\colhead{F160W}
}
\startdata
LACES104037 &
$24.30^{+0.11}_{-0.10}$ &
$24.20^{+0.16}_{-0.14}$ &
$24.24^{+0.17}_{-0.15}$ &
$24.35^{+0.18}_{-0.16}$ &
$24.17^{+0.16}_{-0.15}$  
\\
LACES104037s &
$26.58^{+0.07}_{-0.07}$ &
$25.82^{+0.04}_{-0.04}$ &
$25.89^{+0.06}_{-0.06}$ &
$25.88^{+0.09}_{-0.08}$ &
$24.94^{+0.08}_{-0.08}$ 
\\
LACES94460\tablenotemark{a} &
--- &
--- &
--- &
--- &
$26.12^{+0.09}_{-0.08}$ \\
\enddata

\tablecomments{
All magnitudes are reported in the AB system.Uncertainties correspond to the 16th and 84th percentiles of 10,000 Monte Carlo realizations based on the measured flux uncertainties.
}

\tablenotetext{a}{
Only F160W imaging is available for this source.
}

\end{deluxetable}

\begin{deluxetable}{lcc}
\tablewidth{0.8\textwidth}
\tablecaption{Physical properties \label{tab:laces104037s_and_laces9_phys_all}}
\tablehead{
\colhead{Parameter} &
\colhead{LACES104037s}&
\colhead{LACES94460}
}
\startdata
$E(B\!-\!V)$\tablenotemark{a} &
$0.56^{+0.13}_{-0.11}$ &
$0.024^{+0.05}_{-0.02}$ \\
$12+\log(\mathrm{O/H})$\tablenotemark{b} &
$8.108\pm0.014$ &
$8.318\pm 0.004$ \\
$\log(U)$\tablenotemark{b} &
$-2.58\pm 0.1$ &
$-1.02\pm 0.1$ \\
$\log_{10}(M_{\star}/M_{\odot})$\tablenotemark{b} &
$9.04 ^{+0.01}_{-0.01}$ &
$7.69 ^{+0.02}_{-0.02}$ \\
SFR\tablenotemark{c} ($M_{\odot}\,\mathrm{yr}^{-1}$) &
$83 \pm 40$ &
$6 \pm 3$ \\
\enddata
\tablecomments{All reported uncertainties represent 1$\sigma$ errors.}
\tablenotetext{a}{Derived using \texttt{PyNeb} based on emission-line diagnostics.}
\tablenotetext{b}{Derived from SED fitting.}
\tablenotetext{c}{Estimated following \cite{Kennicutt_1998}.}
\end{deluxetable}

\end{document}